\documentclass[acmsmall,screen]{acmart}
\AtBeginDocument{%
  }
\usepackage{adjustbox}
\usepackage{mdframed}
\setcopyright{acmlicensed}
\copyrightyear{2026}
\acmYear{2026}
\acmDOI{XXXXXXX.XXXXXXX}

\begin{document}

\title{Does Order Matter? An Empirical Investigation into the Impact of File Ordering on Code Review Effectiveness}

\author{Md Shamimur Rahman}
\email{shamimur.rahman@usask.ca}
\orcid{https://orcid.org/0009-0001-5355-4600}
\affiliation{%
  \institution{University of Saskatchewan}
  \city{Saskatoon}
  \state{Saskatchewan}
  \country{Canada}
}

\author{Zadia Codabux}
\affiliation{%
  \institution{University of Saskatchewan}
  \city{Saskatoon}
  \state{Saskatchewan}
  \country{Canada}}
  \email{zadiacodabux@ieee.org}

\author{Chanchal K. Roy}
\affiliation{%
  \institution{University of Saskatchewan}
  \city{Saskatoon}
  \state{Saskatchewan}
  \country{Canada}}
  \email{chanchal.roy@usask.ca}

\renewcommand{\shortauthors}{Rahman et al.}

\begin{abstract}
Modern code review is central to quality assurance in software development practice, yet its effectiveness depends not only on reviewer expertise and change characteristics, but also on how review tools present code changes. Most code review platforms display modified files alphabetically by default, although our prior work shows that developers often perceive this ordering as cognitively misaligned with how they understand multi-file pull requests. However, it remains unclear whether such ordering-related attention patterns are associated with measurable review outcomes at scale. This article presents a large-scale empirical study of file ordering and review effectiveness. We mine 330,343 multi-file pull requests comprising 756,814 file instances from 182 GitHub projects across five programming languages. We examine whether a file's position within a pull request is associated with later involvement in bug-fixing changes, whether pull request size moderates this relationship, and whether reviewer attention, proxied by review comments, aligns with latent bug outcomes. Our results show statistically significant but modest associations between file position, pull request size, review activity, and latent bug likelihood. Later-positioned files exhibit slightly higher latent bug rates, increasing from 56.7\% at position 1 to 61.5\% at position 30. Pull request size shows a non-linear relationship with latent bug likelihood: mid-sized pull requests around ten files have the lowest observed risk, while both very small and very large pull requests show elevated rates. Hurdle models further show that review attention is diluted as pull request size grows, with each additional modified file reducing the odds that a file receives any review comment by approximately 8.7\%. Together, these findings reveal an attention-effectiveness gap: visible review activity does not necessarily translate into defect prevention. Our results suggest that alphabetical ordering should not be treated as a neutral interface default, but as a structural feature that can shape attention allocation, review coverage, and confidence in review outcomes. These findings motivate concrete design directions for review tooling, including context-aware file ordering, dependency-aware grouping, risk-aware prioritization, and per-file coverage indicators that make review attention more visible and actionable.
\end{abstract}



\received{20 February 2007}
\received[revised]{12 March 2009}
\received[accepted]{5 June 2009}

\maketitle

\section{Introduction}
Peer code review is a widely adopted practice in software development, where code authored by a developer is examined by peers before integration into the main codebase. This collaborative process plays a critical role in maintaining software quality by enabling early detection of defects, technical debt, and code smells, as well as team-wide knowledge sharing and adherence to coding standards and project-specific guidelines \cite{tao2015partitioning, bavota2015four, barnett2015helping, wang2015comparative, bosu2013impact, bosu2016process, morales2015code, mcintosh2016empirical}. Leading tech companies such as Microsoft \cite{bacchelli2013expectations}, Google \cite{sadowski2018modern}, and Oracle \cite{cohen2006best}, as well as prominent open-source projects like Android, Qt, OpenStack \cite{mukadam2013gerrit}, and Eclipse \cite{loeliger2012version}, have made code review a core part of their development workflows. While early code reviews were conducted through in-person or checklist-driven inspections \cite{fagan2002design}, Modern Code Review (MCR) practices leverage collaborative tools \cite{mukadam2013gerrit, bosu2012peer, tsotsis2011meet, blischak2016quick} that support asynchronous, distributed collaboration across geographically dispersed teams \cite{bacchelli2013expectations}. In MCR, developers submit Pull Requests (PRs), logical sets of changes that introduce new features, enhancements, or bug fixes, which are then evaluated by experienced developers or maintainers. Popular tools like GitHub \cite{blischak2016quick} and Gerrit \cite{mukadam2013gerrit} typically present modified files in alphabetical order by file path. For example, a PR modifying \textit{src/auth/login.js}, \textit{core/utils.js}, and \textit{src/ui/button.js} would display them in the order: \textit{core/utils.js}, \textit{src/auth/login.js}, \textit{src/ui/button.js}.

Empirical evidence shows that review outcomes are shaped not only by the technical attributes (e.g., size, complexity, ownership) of a change but also by human factors such as attention limits, cognitive load, fatigue, and workflow constraints imposed by tooling \cite{sadowski2018modern, davila2021systematic}. Moreover, the review process changes how reviewers inspect the set of modified files within a PR. In widely used platforms such as GitHub \cite{blischak2016quick}, and Gerrit \cite{mukadam2013gerrit}, these files are presented to reviewers in alphabetical order by default \cite{baum2017optimal, fregnan2022first, olewicki2024empirical}. This default alphabetical ordering is a design choice favoring predictability and implementation simplicity over contextual relevance \cite{bouraffa2025not}. Prior research has firmly established that the position of a file within a PR significantly influences reviewer behavior, i.e., files listed earlier in the sequence receive disproportionately more attention and comments than those listed later, a phenomenon known as positional bias \cite{bagirov2023assessing, fregnan2022first}. In a controlled experiment with 106 participants, Fregnan et al. \cite{fregnan2022first} showed that bugs in the first-positioned file were 64\% more likely to be detected than those in the last-positioned file, and reviewers spent significantly less time on later files. Baum et al. \cite{baum2017optimal} also confirmed that reviewers tend to start from the top of the file list, leading to an uneven distribution of attention. Bouraffa et al. \cite{bouraffa2025not} further observed that reviewer commenting sequences often deviate from the default alphabetical sequence, suggesting that the default order frequently misaligns with reviewers' natural inspection preferences. These studies establish that file ordering influences reviewer behavior. However, they leave open a more consequential question: \textit{Does this ordering-related attention pattern correspond to measurable review effectiveness at scale?}

To ground this question in developer experience, we previously conducted a large-scale survey of 1,355 professional developers across 67 countries and 182 open-source projects~\cite{rahman2026icse}. That study (henceforth the \textit{motivating study}) revealed striking misalignment between tool defaults and developer cognition: only 10.2\% considered alphabetical ordering optimal, more than half (57.6\%) reported that it increases context switching and contributes to review fatigue, and 63.9\% expressed concern that the default ordering may cause them to miss bugs. Developers strongly desired dependency-aware grouping and customizable ordering (requested by 66\% of reviewers). While these findings demonstrate that alphabetical ordering is perceived as cognitively misaligned, they do not establish whether positional bias is associated with review outcomes in project history. This distinction is important because review activity and review effectiveness are not equivalent. Prior code review studies often use review comments as an observable signal of reviewer attention or engagement \cite{mcintosh2016empirical,fregnan2022first,bagirov2023assessing}. However, comments do not necessarily indicate that a defect has been prevented. A file may receive many comments because it is complex, controversial, poorly written, or simply positioned early enough to attract more attention \cite{baum2017optimal,fregnan2022first,bagirov2023assessing}. Conversely, a file may receive no comments because it is correct, because it was inspected silently, because it was overlooked, or because reviewer fatigue and cognitive-load constraints reduced scrutiny in later stages of the review \cite{fregnan2022first,baum2019cognitive,mohanani2018cognitive,spadini2020primers}. Therefore, comment volume should be treated as an observable but incomplete proxy for review attention, not as a direct measure of defect prevention. To determine whether file ordering matters for software quality, this study examines whether file position and reviewer attention are associated with later bug-fixing activity, a commonly used historical proxy for post-release or post-review defect involvement. \cite{mockus2000identifying, mcintosh2016empirical, islam2017comparative}.

To address this gap, this study presents the first large-scale empirical investigation of whether file position within a PR and PR size predict the likelihood of latent bugs, defined as defects that escape review and are subsequently identified in downstream bug-fixing commits. We mined 330,343 multi-file PRs comprising 756,814 file instances from 182 GitHub projects spanning Java, Python, C, C++, and JavaScript, representing the most comprehensive investigation of file-position effects on review effectiveness to date. Our findings confirm and quantify the concerns expressed by developers in the motivating survey study \cite{rahman2026icse}. First, later-positioned files exhibit higher latent bug rates, suggesting that the order in which files are presented can shape review effectiveness beyond merely influencing navigation behavior and represent a persistent pattern in how defects escape review across large-scale project histories. Second, PR size exhibits a non-linear relationship with latent bug likelihood. Very small and very large PRs show elevated latent bug rates, while mid-sized PRs, around ten files, show the lowest observed risk. This finding challenges the common expectation that smaller PRs are inherently safer, an assumption motivated by prior evidence and review guidelines suggesting that smaller changes are easier to inspect and that larger reviews can overwhelm reviewers and reduce defect discovery~\cite{cohen2006best,rigby2013convergent}. Very small PRs may lack sufficient context for reviewers to evaluate the change fully, whereas large PRs may increase cognitive burden and dilute attention across files. Third, our hurdle models show that review attention decreases as PR size increases. Each additional modified file reduces the odds that an individual file receives any review comment by approximately 8.7\%, and also reduces the expected number of comments among files that do receive comments. 

Together, these findings reveal what we call an \emph{attention-effectiveness gap}. Files that appear earlier or receive comments attract observable review activity, but this activity does not necessarily translate into lower latent bug involvement. This gap matters because review tools and development teams often treat visible comments as a signal of review effort and assurance. Our results suggest that such assurance may be unevenly distributed across files, partly shaped by tool-imposed ordering and PR scope. This highlights the need for review interfaces that help reviewers allocate attention more deliberately across all modified files, rather than relying on default ordering and visible comments as implicit indicators of review coverage. 

Therefore, the contributions of this paper are as follows:

\begin{itemize}
\item \textbf{Motivating evidence from practitioners:} We summarize key findings from our prior large-scale survey~\cite{rahman2026icse} of 1,355 developers that establishes the perception gap around alphabetical file ordering and motivates the empirical investigation.

\item \textbf{First large-scale empirical study linking file position to latent defects:} We analyze 330K PRs and 756K file revisions from 182 GitHub projects across five languages to establish whether file position and PR size predict post-review bug likelihood.

\item \textbf{Characterization of the attention--effectiveness gap:} We identify and quantify the counter-intuitive pattern in which files receiving more reviewer comments are not less likely to be implicated in subsequent bug fixes, demonstrating that review activity does not equal defect prevention.

\item \textbf{Evidence of PR-size-modulated review vulnerability:} We show that latent defect risk follows a non-linear pattern across PR size, with mid-sized PRs (around 10 files) exhibiting the lowest risk and both very small and very large PRs showing elevated rates.

\item \textbf{Open replication package:} We release a language-diverse dataset, reproducible methodology, and replication package\footnote{\url{https://doi.org/10.5281/zenodo.20453261}} to enable future research.
\end{itemize}

\section{Background and Related Work}
This section provides an overview of the contextual background and reviews relevant literature across the following key areas.

\subsection{Cognitive Constraints in Code Review}
While reviewers support software extensibility and defect mitigation, the effectiveness of code review is limited by cognitive constraints and biases that hinder the accurate evaluation of code changes \cite{bacchelli2013expectations, sadowski2018modern}. Cognitive load theory suggests that review tasks impose substantial mental demands, including sustained attention, mental model construction, reasoning about delocalized bugs, and comprehension of large or multi-file changes, often exceeding working memory capacity \cite{baum2017optimal, baum2019associating, gonccalves2020explicit, gonccalves2022explicit}. Beyond general cognitive load, other cognitive constraints, including confirmation bias, anchoring effects, availability bias, and representativeness heuristics, could deviate reviewers from rational judgment processes and compromise review effectiveness \cite{chattopadhyay2020tale, huang2020biases, spadini2020primers, thongtanunam2020review}. For instance, Huang et al. \cite{huang2020biases} conducted an eye-tracking and medical imaging study to explore how biases manifest during the inspection process, revealing that developers often rely on familiar patterns or recent experiences when assessing unfamiliar code. Moreover, attention switching, decision fatigue (e.g., inattention, impulsivity, procrastination), and a mismatch between reviewer expertise and code further lead to slower reviews and degrade review effectiveness \cite{gonccalves2022explicit, jetzen2025towards}.

\subsection{Psychology of File Position in PR Review}
Human cognitive and psychological limitations influence the thoroughness of code reviews, particularly in how attention is distributed across files within a PR. Baum et al. \cite{baum2017optimal} found that reviewers tend to begin at the top of the file list, leading to uneven attention. In a large-scale study, Fregnan et al. \cite{fregnan2022first} demonstrated that top-positioned files receive more comments. In a controlled experiment involving 106 participants, bugs in the first file were 64\% more likely to be detected than those in the last. Bagirov et al. \cite{bagirov2023assessing} further confirmed that reviewers focus disproportionately on earlier files in closed-source industrial projects. These findings highlight how attention diminishes over the review sequence, a phenomenon called attention decrement, where sustained focus declines over time, particularly with sequential tasks \cite{murdock1962serial, vandenbos2007apa, hendrick1973attention}. This decay introduces positional bias, challenging the assumption that all files receive equal scrutiny. Such uneven attention distribution can influence review effectiveness, for instance, McIntosh et al. \cite{mcintosh2016empirical} showed that differences in review coverage correlate with software quality, highlighting the practical consequences of reviewer attention imbalance. Another key psychological factor influencing review performance is the limited capacity of human working memory, which involves processing and manipulating information during complex tasks \cite{baddeley2003working, baum2019associating, wilhelm2013working}. While individual capacity varies, research consistently shows that working memory can only maintain a limited amount of information \cite{cowan2001magical, cowan2010magical}. This limitation plays a key role in cognitively demanding programming tasks, such as understanding deeply nested logic or tracking non-local dependencies \cite{bergersen2011programming}. Moreover, in multi-file PRs, as cognitive load accumulates, reviewers may struggle to retain context across files, reducing attention and comprehension as the review progresses.

\subsection{Investigation on Reviewing Code Changes}
To decide whether to accept, reject, or request changes to a PR, reviewers must understand code modifications in context and assess their necessity and quality. Baum et al. \cite{baum2017optimal} found that reviewers often follow the default file order, typically alphabetical by file path, though later work suggests more effective strategies. Gonccalves et al. \cite{gonccalves2022explicit} recommend grouping related changes, while Bagirov et al. \cite{bagirov2023assessing} showed that ordering files by change size surfaces critical or error-prone components earlier than alphabetical sorting. Bouraffa et al. \cite{bouraffa2025not} found that 44.6\% of reviewers deviated from alphabetical order, often starting with large diffs, files tied to the PR description, or tests when both test and production code are present. Olewicki et al. \cite{olewicki2024empirical} proposed a similarity-based ordering to highlight files needing attention, though navigation patterns varied with reviewer familiarity and preferences. Beyond file order, eye-tracking research reveals that certain code elements inherently attract reviewer focus independent of their location. Abid et al. \cite{abid2019developer} found that developers primarily focus on function calls, followed by control flow constructs and method declarations, consistent with Rodeghero et al. \cite{rodeghero2015eye}, who identified method signatures as the primary focal point for reviewers, with substantial attention also given to call terms and control structures. Additionally, Al Madi et al. \cite{al2021novice} showed that tokens with low frequency and higher character length tend to hold reviewers’ attention longer, suggesting that both semantic relevance and lexical complexity influence visual engagement.

\subsection{Defect Detection in Code Review}

Code review is a vital quality assurance practice for detecting defects that often go undetected by traditional testing. Empirical studies demonstrate that it excels at identifying evolvability defects that hinder future development without compromising immediate functionality. Siy et al. \cite{siy2001does} found that 75\% of defects detected via code review fall into this category, unlike execution-based testing, which targets functional defects. The studies \cite{mantyla2008types, yu2023security} further confirmed that code reviews emphasize maintainability-related issues, positioning code review as a complement to traditional testing in long-term maintenance contexts. However, the cognitively demanding and time-consuming nature of code review often leads to reduced attention toward later parts of a change-set, increasing the risk of undetected defects \cite{fregnan2022first, bagirov2023assessing, yu2023security}. While automated tools help ease the burden, human reviewers remain essential for identifying complex, context-dependent issues beyond current tool capabilities \cite{yu2023security, tufano2024code, tufano2025deep}. To support manual review, researchers proposed risk assessment frameworks based on historical patterns of defects. The studies \cite{goccmen2025enhanced, thongtanunam2015investigating} suggested measuring file-level bug frequency from prior bug-fixing PRs as an indicator of modification risk. They also observed that larger PRs, as indicated by line changes or file count, are associated with lower review effectiveness. These findings imply that file position within a PR may influence defect detection, though empirical exploration of this relationship remains limited. 

While prior work demonstrates that file position affects reviewer attention and navigation behavior, its relationship with downstream software quality remains underexplored. Existing studies show that earlier files tend to receive more attention and that reviewers often deviate from default alphabetical ordering, but they do not establish whether files appearing at different positions are more or less likely to be involved in later bug-fixing changes. This study addresses that gap by empirically examining the relationship between file order in multi-file PRs, reviewer attention, PR size, and subsequent bug-fixing activity across 182 open-source projects. In doing so, we move beyond behavioral and perceptual evidence to investigate whether positional bias represents a measurable review-effectiveness concern. Our findings position file ordering as an important but previously underexamined factor in code review quality, with practical implications for review tool design, attention allocation, and defect management.

\section{Motivating Study: Developer Perceptions of File Ordering}
\label{sec:motivating}
The repository-mining study in this paper is motivated by our prior ICSE 2026 study on developers' perceptions of file ordering in code review~\cite{rahman2026icse}. That study did not examine downstream defect outcomes. Instead, it investigated whether the default alphabetical ordering used by code review tools aligns with how developers actually navigate and reason about multi-file PRs. We summarize it here to establish the practitioner-facing problem that motivates the present empirical investigation.

\subsection{Overview}

The study conducted a large-scale online survey of 1,355 professional developers from 67 countries and 182 open-source projects. Participants were recruited from active Apache Software Foundation and highly popular GitHub repositories using publicly available contact information, supplemented by snowball sampling. The survey focused on developers with code review experience and asked how they navigate multi-file PRs, whether they follow or deviate from default alphabetical ordering, how they perceive alphabetical ordering in terms of efficiency and cognitive fit, what challenges they encounter in multi-file reviews, and what tool support they expect for improving file navigation.

The study used a mixed-methods design. Closed-ended responses were 
analyzed using descriptive statistics, statistical significance tests, effect-size analysis, and regression models. Open-ended responses were analyzed through manual coding and thematic analysis by two independent evaluators, with inter-rater agreement exceeding Cohen's $\kappa >0.80$. The purpose of the survey was not to establish whether alphabetical ordering causes defects, but to determine whether practitioners perceive file ordering as a meaningful review problem and whether their experiences justify a deeper outcome-based investigation. The survey instrument, analysis scripts, and supplementary materials are publicly available in the replication package~\cite{rahman2026icse}.

\subsection{Key Findings}

The survey produced several findings that are directly relevant to file ordering, cognitive burden, and perceived review effectiveness. Together, they reveal a consistent gap between how review tools present changed files and how developers naturally reason about multi-file code changes.

\begin{itemize}
    \item \textbf{Alphabetical ordering is widely perceived as suboptimal.} 
    Only 10.2\% of respondents considered alphabetical ordering optimal for navigating multi-file PRs. Many developers instead preferred strategies based on logical dependency, semantic relatedness, change impact, file type, or the relationship between production and test files.

    \item \textbf{Developers frequently work around the default order.} 
    Although alphabetical ordering is the default in common review tools, many reviewers reported adopting alternative navigation strategies, such as starting with core logic, reviewing tests first, prioritizing larger or riskier diffs, following the PR description, or mentally regrouping related files.

    \item \textbf{Alphabetical ordering increases perceived cognitive burden.} 
    More than half of respondents (57.6\%) reported that alphabetical ordering increases context switching, disrupts logical reasoning across related files, and contributes to review fatigue, especially when PRs contain many files or scattered changes. Developers also described difficulty tracking cross-file dependencies, maintaining contextual coherence, and understanding the broader impact of changes when related files are separated by the default order.

    \item \textbf{Developers associate file order with missed defects.} 
    63.9\% of respondents expressed concern or uncertainty that default file ordering may cause them to overlook bugs. This finding provides practitioner-level motivation for examining whether file position is empirically associated with later bug-fixing activity.

    \item \textbf{Reviewers want more flexible and context-aware ordering support.} 
    Participants expressed strong demand for dependency-aware grouping and customizable file ordering, requested by 66\% of reviewers, as well as presentations that better reflect semantic relationships among changed files.
\end{itemize}

\subsection{Motivation for the Present Study}

The motivating study establishes that alphabetical file ordering is not merely a minor usability preference. Many developers perceive the default order as misaligned with how they understand multi-file changes, particularly when related files are separated, critical files appear late in the list, or reviewers must mentally reconstruct the logical flow of a change~\cite{rahman2026icse}. These perceptions suggest that file ordering may influence not only navigation convenience, but also the cognitive conditions under which reviewers allocate attention and assess correctness~\cite{fregnan2022first,bagirov2023assessing,rahman2026icse}. However, our published ICSE study was intentionally perception-focused. Its findings are based on developers' self-reported experiences, preferences, and concerns~\cite{rahman2026icse}. As such, they cannot determine whether the perceived problems translate into measurable review outcomes in project history. In particular, the survey leaves three outcome-oriented questions unanswered. First, are files appearing later in a PR more likely to be involved in subsequent bug-fixing changes? Second, does PR size amplify or moderate the relationship between file position and latent bug likelihood? Third, does visible review activity, such as review comments, correspond to lower post-review defect involvement, or can comments reflect attention without necessarily indicating effective defect prevention?

The present study addresses these questions through large-scale repository mining. Rather than re-examining whether developers dislike alphabetical ordering, we investigate whether file position, PR size, and reviewer attention are empirically associated with latent bug likelihood across 182 open-source projects. In this way, the prior study serves as motivating and triangulating evidence, while the core contribution of this study is an outcome-based analysis of review effectiveness. This positioning allows us to connect developer perceptions with observable repository evidence, moving from the question of whether alphabetical ordering feels cognitively misaligned to whether ordering-related patterns are associated with downstream software quality. For transparency, Table~\ref{tab:icse tosem difference} summarizes the intended distinction.

\begin{table}[htbp]
\centering
\small
\caption{Positioning the current study relative to our prior survey paper.}
\label{tab:icse tosem difference}
\begin{adjustbox}{width=\linewidth}
\begin{tabular}{p{0.18\textwidth} p{0.35\textwidth} p{0.40\textwidth}}
\toprule
\textbf{Dimension} & \textbf{Motivating Survey Study ~\cite{rahman2026icse}} & \textbf{This Study} \\
\midrule
Main focus
  & Practitioner perceptions of alphabetical file ordering
  & Empirical association between file order, PR size, review attention, and latent defects. \\\cmidrule{2-3}

Primary evidence
  & Survey of 1,355 developers
  & Repository mining of 330,343 PRs and 756,814 file instances \\\cmidrule{2-3}
Unit of analysis
  & Developer/reviewer response
  & File instance within a multi-file PR \\\cmidrule{2-3}
Primary outcome
  & Perceived optimality, cognitive burden, and ordering preferences
  & Later involvement in bug-fixing changes as a latent bug proxy \\\cmidrule{2-3}
Role in this study
  & Motivation and triangulation
  & Core empirical contribution \\\cmidrule{2-3}
Main contribution
  & Shows that alphabetical ordering is perceived as cognitively misaligned
  & Shows that ordering and PR scope are associated with measurable review coverage and latent-defect patterns \\
\bottomrule
\end{tabular}
\end{adjustbox}
\end{table}

\section{Study Methodology} \label{sec:design} 
Motivated by the practitioner concerns established in Section~\ref{sec:motivating}, we now present our empirical study. We employ a large-scale historical software repository mining approach combined with statistical analysis to examine whether file-position and PR-size effects on review behavior translate into measurable differences in post-review defect likelihood. 

\subsection{Research Questions}
To align the Research Questions (RQs) with the goal of this study, we 
decompose the review effectiveness into three related aspects that connect tool-imposed file ordering to downstream defect outcomes. First, if alphabetical ordering creates positional bias, the most direct outcome-level question is whether a file's position within a PR is associated with later involvement in bug-fixing changes. This motivates RQ1, which examines file position as the primary ordering-related factor. Second, positional effects may not occur uniformly across all PRs. Reviewing two or three files is cognitively different from reviewing dozens, where attention, context retention, and fatigue become more salient~\cite{cowan2001magical, baum2019cognitive}. PR size is therefore a necessary contextual factor for understanding when positional effects are likely to matter, motivating RQ2. Third, prior work commonly uses review comments as an observable signal of reviewer attention~\cite{mcintosh2016empirical, fregnan2022first, 
bagirov2023assessing}, but attention is not necessarily equivalent to defect prevention. To determine whether visible review activity aligns with downstream quality outcomes, we examine the relationship between comment activity and latent bug likelihood, motivating RQ3.

Therefore, the three RQs operationalize a progression from interface-level factor (file position) to review-context factor (PR size) to behavioral mechanism (reviewer attention), allowing us to establish not only whether ordering-related patterns are associated with latent defects, but under what conditions this association is most pronounced, and whether observable review activity can account for it.

\begin{itemize}
    \item \textbf{RQ1: To what extent is a file's relative position within a multi-file PR associated with its likelihood of being involved in a subsequent bug-fixing change?} This RQ examines the direct outcome-level relationship between file ordering and the likelihood of latent bugs. Since alphabetical ordering determines the sequence in which reviewers encounter files, RQ1 tests whether files that appear earlier or later in that sequence differ in their subsequent involvement in bug fixes.

    \item \textbf{RQ2: How does PR size moderate the relationship between file position and latent bug likelihood?} This RQ examines whether positional effects depend on the review context. Larger PRs require reviewers to inspect more files, maintain more cross-file context, and distribute attention across a broader change set. Therefore, PR size may amplify, weaken, or reshape the relationship between file position and the likelihood of latent bug.

    \item \textbf{RQ3: Does reviewer attention, proxied by review comment activity, align with latent bug outcomes, and how is this relationship affected by PR size?} This RQ examines whether observable review activity corresponds to effective defect prevention. If files receiving more comments are still frequently involved in subsequent bug-fixing changes, this would suggest that review activity and review effectiveness are not equivalent, and that attention may be diluted as PR size increases.
\end{itemize}

\subsection{Subject Projects}

We begin by investigating active and popular open-source projects hosted by the Apache Software Foundation\footnote{\url{https://projects.apache.org/}} on GitHub. Of its 481 active projects across 36 languages, the majority (262) are implemented in Java. Following common MSR practice, we use repository popularity, maturity, community size, and recent development activity as project-selection criteria, since these signals help filter out toy, inactive, or weakly maintained repositories and increase the likelihood of observing meaningful review activity~\cite{kalliamvakou2014promises, gousios2014dataset,borges2018s,elazhary2019not}. Specifically, we select 133 projects that meet the following inclusion criteria: i) at least 500 stars, ii) a minimum of 5,000 commits, and iii) active open PRs to ensure recent activity. This filtering targets projects with strong activity and community engagement.

Since Apache projects are maintained under a single foundation and contributors may overlap across Java projects, we expand our dataset to improve generalizability by collecting 49 additional projects from GitHub using the SEART\footnote{\url{https://seart-ghs.si.usi.ch/}} search engine. The two project groups use different thresholds because they serve complementary sampling purposes. Apache projects are drawn from a curated foundation-managed ecosystem, where projects already satisfy organizational maturity, governance, and maintenance expectations. Therefore, moderate thresholds on stars, commits, and active pull requests are sufficient to identify projects with meaningful review activity. In contrast, the broader GitHub ecosystem contains a much larger and noisier population of repositories, including personal, experimental, inactive, or weakly maintained projects. For this second group, we apply stricter thresholds to ensure that the selected repositories are mature, popular, community-driven, and likely to contain sustained code review activity. Specifically, these projects meet the following criteria: at least 10,000 commits, 100 contributors, 10,000 stars, and active open PRs. Commits indicate development maturity, contributors approximate community breadth, stars are commonly used as a popularity signal in GitHub-based empirical studies, and active PRs indicate ongoing collaborative development~\cite{borges2018s,elazhary2019not}. Unlike the Apache dataset, which is Java-dominant, these additional projects span multiple languages, including Java, Python, C, C++, and JavaScript. Although some Java-based Apache projects could appear in both the Apache and additional GitHub samples, we explicitly checked for and removed overlapping repositories before conducting the analysis.

\subsection{Data Collection} After selecting the projects, we collected only the \textit{``merged”} PRs using the GitHub REST API\footnote{\url{https://docs.github.com/en/rest}}. This yielded 256,824 PRs from 133 Apache projects and 682,187 PRs from 49 additional projects, totaling 939,011 PRs across 182 projects. Of these, 43.26\% (406,266 PRs) had one or more reviewers. To analyze review patterns based on file count, we categorized PRs as follows: 17.45\% (75,923) changed one file, 39.06\% (169,954) changed two to five files, 14.01\% (60,964) changed six to nine files, and 24.48\% (99,425) changed ten or more files, with a maximum of 73 files. We excluded single-file PRs, as they offer limited insight into reviewer navigation and file prioritization. For each PR, we collected metadata including author, reviewers, review comments, commits, commit messages, file-level changes per commit, and associated code diffs.

\subsection{Data Preparation}\label{Data Preparation} 
To answer $\mathbf{RQs}$, we performed several preprocessing steps to structure the collected PRs information. 

\subsubsection{Bug-fixing PRs}
To examine the correlation with software defects, we first identify PRs containing bug-fixing commits. Following the established heuristic approach proposed by Mockus et al. \cite{mockus2000identifying}, we apply keyword-based filtering using six specific keywords (`bug,' `fix,' `fixup,' `error,' `crash,' and `fail') within commit messages to detect commits addressing prior defects. These keywords reflect how developers typically document bug-related changes, and prior studies confirm their effectiveness in identifying bug-fixing commits \cite{barbour2013empirical, islam2017comparative}. Since a PR may include multiple commits, we label it as bug-fixing if at least one commit matches the criteria. Moreover, we manually validated this heuristic by randomly sampling 384 commits (95\% confidence interval, 5\% margin of error). Two independent evaluators (more than 12 years of software development experience) achieved strong inter-rater agreement (Cohen's $\kappa$ = 0.97) \cite{cohen1960coefficient}, with 96.61\% and 99.18\% individual alignment with our labels, confirming the heuristic's precision.

\begin{figure}[htbp]
\vspace{-1.0em}
\centering     
\includegraphics[width=8.0cm]{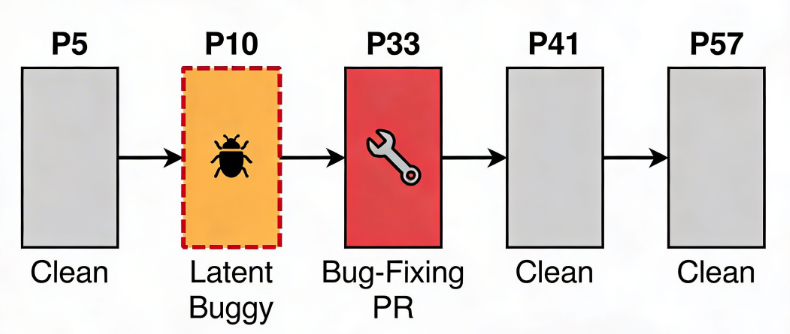}
\caption{Identifying latent bug files using sequential PR analysis.}
\vspace{-0.5em}
\label{latenbug}
\end{figure}

\subsubsection{Latent Bug Files.} We identify latent bug files, those with defects undetected in prior reviews, by analyzing the historical sequence of changes for each file within a repository. For every unique file, we extract an ordered list of PRs in which it was modified or reviewed. Figure~\ref{latenbug} illustrates the labeling procedure. For example, if \textit{sampleFile.java} appeared in PRs $P_5$, $P_{10}$, $P_{33}$, $P_{41}$, and $P_{57}$. We consider $P_{33}$ a bug-fixing PR and mark the file as latent buggy if a bug-fixing commit in $P_{33}$ modified the file, implying that the defect was likely introduced or missed in $P_{10}$. The underlying assumption is that when a file reappears in a bug-fixing PR immediately after a prior modification, the earlier review likely failed to detect the defect. Formally, if a file appears in two consecutive PRs ($P_M$, $P_N$), and $P_N$ is a bug-fixing PR, we label the file as latent buggy. Files not followed by such PRs are labeled clean. We also record each file’s relative position within the earlier PR $P_M$ (i.e., index in the file list as presented by the review tool) to support positional analysis. This heuristic is not claimed as ground truth, even if a subsequent modification is not strictly a bug fix, its necessity often indicates that the earlier review failed to catch omissions or areas requiring refinement \cite{mockus2000identifying}, which aligns with our goal of measuring review effectiveness.

\subsection{Statistical Analysis}

\subsubsection{Exploratory Analysis.} For each file in our dataset, we recorded four attributes: its relative position within the originating PR, the PR size measured as the number of modified files, its latent bug status, and the number of reviewer comments it received. To enable stable descriptive comparisons across PR contexts, we grouped PR size into five bands: 2--3, 4--5, 6--9, 10--19, and 20+ 
files. These bands are intentionally unequal in width and reflect qualitative differences in cognitive demand rather than equal numeric 
intervals. Very small PRs of 2--3 files impose minimal context-switching burden and can typically be reviewed in a single pass. PRs with 4--5 files begin to require more deliberate navigation but remain manageable within working memory limits~\cite{cowan2001magical}. PRs of 6--9 files introduce cross-file dependency tracking as a non-trivial cognitive demand, while PRs of 10--19 files are associated with measurably higher reviewer effort and reduced per-file attention in prior work~\cite{mcintosh2014impact, thongtanunam2015investigating}. PRs of 20 or more files represent the upper range where review dilution is most pronounced. This cognitively motivated grouping is consistent with prior studies that use non-uniform PR size categories to capture meaningful behavioral thresholds rather than arbitrary intervals~\cite{rigby2013convergent, baum2017optimal}. Moreover, files appearing at positions beyond the 30th occurred infrequently and would otherwise yield sparse categories, we therefore collapsed all files appearing after position 30 into a single “30+” group to maintain statistical robustness in subsequent analyses.

\subsubsection{Positional and Size Effects over Latent Bug.} To assess positional effects, we used complementary continuous and categorical statistical tests \cite{powers2008statistical}. In the continuous setting, we compared the distribution of file positions between buggy and non-buggy files using both a two-sample $t$-test \cite{cressie1986use} and the Mann--Whitney $U$ test \cite{mcknight2010mannn} as a nonparametric alternative when the normality assumption may not hold. We further quantified the association between position and bug likelihood using Pearson correlation \cite{benesty2009pearson} for linear relationships and Spearman rank correlation \cite{sedgwick2014spearman} for monotonic relationships. To interpret practical significance alongside statistical significance, we report Cohen's $d$ \cite{diener2010cohen} as an effect-size measure for differences in position distributions. In the categorical setting, we discretized file position into quantile-based bins (e.g., deciles) and applied a $\chi^2$ test of independence \cite{mchugh2013chi} to test whether bug occurrence varies by position category. We report Cram\'er's $V$ \cite{cramer1999mathematical} as an effect size for categorical associations and used a trend test \cite{agresti2011categorical} to assess whether bug rates change monotonically across position bins. 

\subsubsection{Hurdle Models for Review Activity.} To analyze how review activity relates to latent bug status and PR size, we modeled reviewer comment allocation using two-stage hurdle models \cite{feng2021comparison}, which are well-suited to code review data where many files receive zero comments. The first stage uses a logistic component to estimate the probability that a file receives any comments (a binary outcome). The second stage uses a Poisson count component to model the number of comments conditional on receiving at least one. Formally, the probability mass at zero is estimated via logistic regression, while $\mathbb{E}[\textit{Comments}\mid \textit{Comments}>0]$ is estimated via Poisson regression \cite{cameron2013regression}. We fit two specifications: a baseline model that includes only buggy status, and an extended model that adds PR size to evaluate potential confounding and moderation effects. This hurdle framework enables us to disentangle whether higher observed attention to certain files reflects meaningful reviewer prioritization or structural artifacts (e.g., position and PR size) that may not translate into improved defect detection.

\section{Results}
Here, we present findings addressing our RQs, providing insights into how file position and PR size influence bug occurrence. 

\subsection{RQ1--File Position vs. Latent Bug Likelihood}
This RQ examines whether a file’s relative position within a PR and the overall PR size are associated with its likelihood of inducing a latent bug, as indicated by subsequent bug-fixing commits. From 330,343 PRs involving two or more files, we identified 756,814 files later modified in the immediately following PRs. Of these, 57.49\% (435,087) appeared in bug-fixing commits within bug-fixing PRs, suggesting they were likely implicated in earlier defects (see Section~\ref{Data Preparation}). The remaining 42.51\% were not linked to immediate bug fixes, either appearing in non-bug-fixing PRs or in non-bug-fixing commits within bug-fixing PRs. File positions ranged from 1 to 30, and PR sizes from 2 to 30, with sizes 21–30 grouped as \textit{``20plus.''}

\begin{table}[htbp]
\centering
\small
\caption{Statistical results for file position and bug occurrence}
\begin{adjustbox}{width=12cm}
\begin{tabular}{llr}
\toprule
\textbf{Metric} & \textbf{Test / Value} & \textbf{Result} \\
\midrule
File Position Mean           & Buggy vs. Non-Buggy       & 8.95 vs. 8.58 \\
Two-sample \textit{t}-test   & $t$                        & 19.10 ($p < 0.001$) \\
Mann–Whitney \textit{U} test & $U$                        & $7.14 \times 10^{10}$ ($p < 0.001$) \\
Pearson Correlation          & $r$                        & 0.022 \\
Spearman Correlation         & $\rho$                     & 0.018 \\
Cohen’s $d$                  & $d$                        & 0.044 \\
Chi-square (Relative Quantile) & $\chi^2$, df, $V$         & 2889.85, 19, 0.062 ($p < 0.001$) \\
Trend Test                   & Statistic                  & $> 1.2 \times 10^8$ ($p < 0.001$) \\
\bottomrule
\end{tabular}
\label{tab:position-stats}
\end{adjustbox}
\end{table}

To examine positional effects, we conducted both continuous and categorical analyses \cite{powers2008statistical} (Table~\ref{tab:position-stats}). On average, buggy files occurred at slightly higher positions within a PR (mean = 8.95) than non-buggy files (mean = 8.58), a difference supported by a two-sample \textit{t}-test \cite{cressie1986use} ($t = 19.10$, $p < 0.001$) and a Mann–Whitney \textit{U} test \cite{mcknight2010mannn} ($U = 7.14 \times 10^{10}$, $p < 0.001$). Pearson \cite{benesty2009pearson} ($r = 0.022$) and Spearman \cite{sedgwick2014spearman} ($\rho = 0.018$) correlations, along with Cohen’s $d = 0.044$ \cite{rosenthal1994parametric}, indicate a very weak effect, yet statistically significant due to the large sample size and high base bug rate imply that even subtle positional trends may carry practical relevance at scale. To assess non-linear patterns, we applied chi-square tests \cite{mchugh2013chi} across four binning strategies (equal-width, quantile, relative quantile, and domain-informed). All yielded statistically significant results ($p < 0.001$), with the relative quantile binning showing the strongest signal ($\chi^2 = 2889.85$, $df = 19$, Cramer's $V = 0.062$) \cite{akoglu2018user}. A trend test \cite{agresti2011categorical} further confirmed a monotonic increase in bug rates across position bins ($> 1.2 \times 10^8$, $p < 0.001$).

\begin{figure}[htbp]
\centering
\begin{minipage}[t]{0.49\linewidth}
  \centering
  \includegraphics[width=\linewidth]{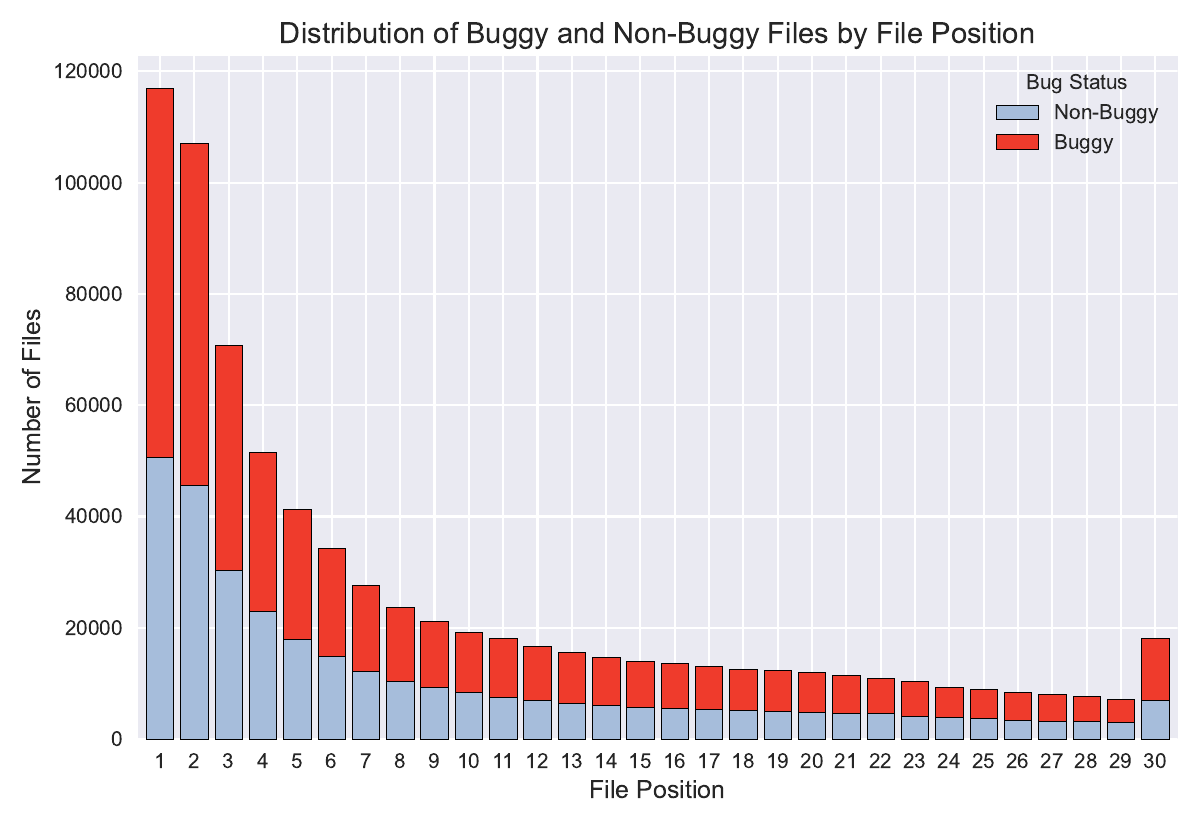}
  \caption{Buggy and non-buggy files ratio by file position.}
  \label{fig:bug-nonbug-ratio}
\end{minipage}\hfill
\begin{minipage}[t]{0.49\linewidth}
  \centering
  \includegraphics[width=\linewidth]{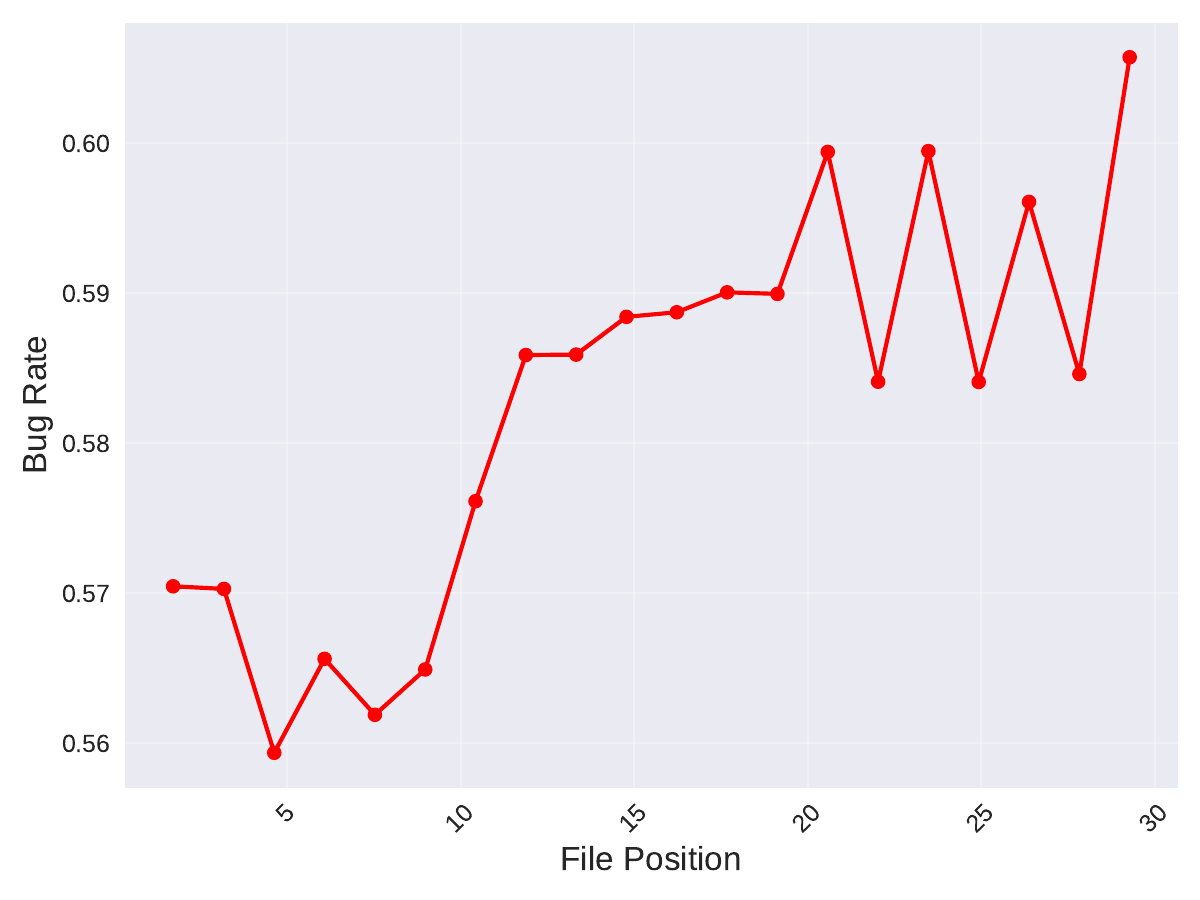}
  \caption{Distribution of bug rate and file position.}
  \label{fig:bugrate-position}
\end{minipage}
\end{figure}

To further investigate the positional prevalence of latent bugs, we conducted one-sided binomial tests \cite{sobel1959group} at each file position (1--30), assessing whether the proportion of buggy files exceeded the 50\% baseline. All positions yielded statistically significant results ($p < 0.001$), with bug rates ranging from 56.7\% at position 1 to 61.5\% at position 30 (Figures \ref{fig:bug-nonbug-ratio} and \ref{fig:bugrate-position}). This consistent overrepresentation of buggy files across positions suggests that latent bugs are pervasive in merged PRs and not confined to particular file locations. Many of these files were subsequently re-modified in the next PR, often as part of bug-fixing commits, indicating that initial code reviews failed to detect these faults. These findings reveal a critical limitation in review effectiveness and underscore the need for more thorough scrutiny across all files in a PR, regardless of their position.

\begin{mdframed}[
    linecolor=black!60,
    linewidth=1.5pt,
    backgroundcolor=yellow!8,
    innertopmargin=6pt,
    innerbottommargin=6pt
]
\noindent\textbf{\underline{RQ1 Findings:}} File position within a PR shows a statistically significant, yet modest, positive association with latent bug introduction. Files appearing at later positions exhibit marginally higher bug rates (56.7\% at position 1 → 61.5\% at position 30), contradicting the hypothesis that earlier-positioned files (which receive more attention) escape undetected defects. However, latent bugs are prevalent at all positions, indicating that the phenomenon is not purely positional and likely reflects additional review and change-related factors.
\end{mdframed}

\subsection{RQ2--PR Size vs. Latent Bug Likelihood}

\begin{figure}[htbp]
\centering     
\includegraphics[width=12cm]{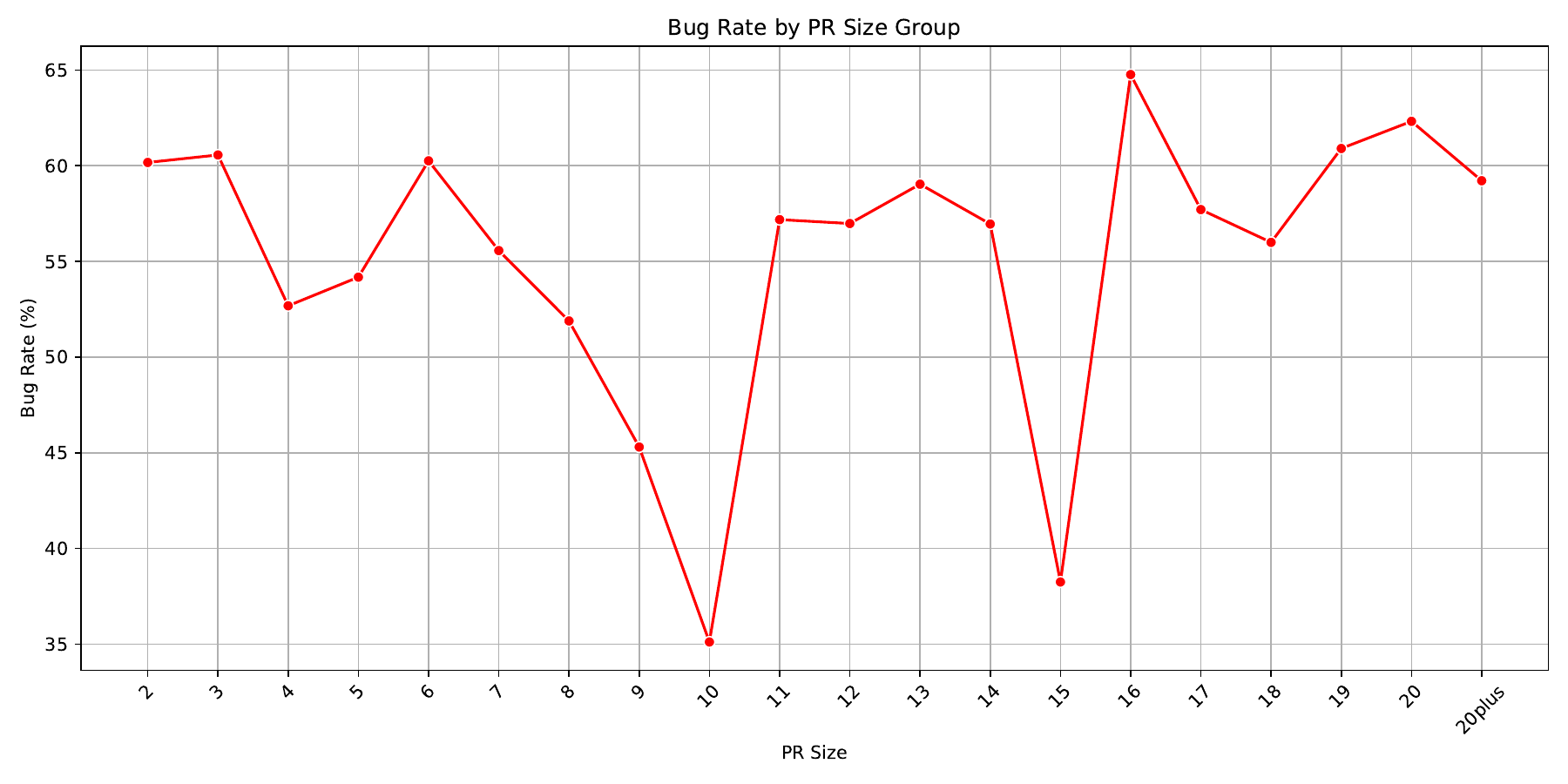}

\caption{Distribution of bugs with respect to pull request size.}
\label{bug rate}
\end{figure}

To assess whether PR size (i.e., number of changed files) itself influences bug propensity (Table~\ref{tab:prsize-stats}), we grouped PR sizes from 2 to 20 and added a ``20plus” category for larger PRs. A chi-square test ($\chi^2 = 6822.32$, $df = 19$, $p < 0.001$) and Kruskal–Wallis test \cite{mckight2010kruskal} ($H = 6822.32$, $p < 0.001$) both revealed significant differences in bug rates across PR size groups. Although the effect size was small (Cramér’s $V = 0.095$), the trend was consistent (Figure \ref{bug rate}), i.e., PRs with 2–3 files showed the highest bug rates ($>$60\%), PRs around size 10 exhibited a dip (e.g., 35.1\% at size 10), and larger PRs (e.g., ``20plus”) showed rates above 59\%. Correlation tests further supported this pattern, with Spearman’s $\rho = 0.017$ and Kendall’s $\tau = 0.015$ \cite{abdi2007kendall} (both $p < 10^{-48}$), indicating a statistically significant but small/modest monotonic relationship. To test robustness, we repeated all analyses on a subset of programming and logic implementation files (filtered by file extension) and observed consistent results, i.e., statistically significant trends with similarly small effect sizes. These findings suggest the overall size of the PR is a statistically reliable, though modest, indicator of future bug involvement.

\begin{table}[htbp]
\centering
\caption{Statistical results for PR size and bug occurrence}
\begin{adjustbox}{width=12cm}
\begin{tabular}{llr}
\toprule
\textbf{Metric} & \textbf{Test / Value} & \textbf{Result} \\
\midrule
Chi-square Test            & $\chi^2$, df              & 6822.32, 19 ($p < 0.001$) \\
Kruskal–Wallis Test        & $H$                       & 6822.32 ($p < 0.001$) \\
Cramér’s $V$               & $V$                       & 0.095 \\
Bug Rate by PR Size        & Size 2–3, 10, 20plus         & $>$60\%, 35.1\%, $>$59\% \\
Spearman Correlation       & $\rho$                    & 0.017  \\
Kendall’s Tau              & $\tau$                    & 0.015  \\
Robustness Check           & File subset analysis      & Consistent trends, small effect sizes \\
\bottomrule
\end{tabular}
\label{tab:prsize-stats}
\end{adjustbox}
\end{table}

\begin{mdframed}[
    linecolor=black!60,
    linewidth=1.5pt,
    backgroundcolor=yellow!8,
    innertopmargin=6pt,
    innerbottommargin=6pt
]
\noindent\textbf{\underline{RQ2 Findings:}} PR size is significantly associated with latent bug likelihood, with the lowest risk concentrated in mid-sized PRs and higher risk at both ends of the size spectrum, indicating that size alone provides a reliable but modest signal of post-review defects.
\end{mdframed}

\begin{table}[htbp]
\centering
\small
\caption{Hurdle models for review comment distribution}
\begin{adjustbox}{width=12cm}
\begin{tabular}{lllrrrr}
\toprule
\textbf{Model} & \textbf{Predictor} & \textbf{Component} & \textbf{Coefficient} & \textbf{OR / IRR} & \textbf{Pseudo-$R^2$} & \textbf{p-value} \\
\midrule
Model 1 & Buggy       & Logistic & 0.1132  & 1.12  & 0.00054 & $<$ 0.001 \\
Model 1 & Buggy       & Poisson  & 0.0564  & 1.058 & 0.00049 & $<$ 0.001 \\
Model 2 & Buggy       & Logistic & 0.1558  & 1.17  & 0.0848  & $<$ 0.001 \\
Model 2 & Buggy       & Poisson  & 0.0605  & 1.062 & 0.0115  & $<$ 0.001 \\
Model 2 & PR Size     & Logistic & $-$0.0915 & 0.91  & --      & $<$ 0.001 \\
Model 2 & PR Size     & Poisson  & $-$0.0174 & 0.983 & --      & $<$ 0.001 \\
\bottomrule
\end{tabular}
\label{tab:hurdle-model}
\end{adjustbox}
\vspace{-0.50em}
\end{table}

\subsection{RQ3--Review Attention}
We employed two Hurdle models \cite{feng2021comparison}, as it handles excess zeros in the dependent variables, to examine how bug status and PR size affect the distribution of reviewer comments. As shown in Table~\ref{tab:hurdle-model}, the initial model, which included only \texttt{bugfixing} as a predictor, buggy files were significantly more likely to receive at least one comment (logistic coefficient = 0.113, $p < 0.001$; odds ratio = 1.12; pseudo-$R^2$ = 0.0005) and, when commented upon, received 5.8\% more comments on average (Poisson coefficient = 0.056, $p < 0.001$; incidence rate ratio = 1.058; pseudo-$R^2$ = 0.0005). Incorporating \texttt{PR\_size} (i.e., pull request size) into the model improved explanatory power (pseudo-$R^2$ = 0.085 for the binary component; 0.012 for the count component) and revealed a strong negative association between PR size and review activity. Each additional file in a PR decreased the likelihood of receiving a comment by 8.7\% (coefficient = $-0.092$, odds ratio = 0.91, $p < 0.001$) and reduced the expected number of comments by 1.7\% (coefficient = $-0.0174$, incidence rate ratio = 0.983, $p < 0.001$). Notably, the inclusion of PR size slightly amplified the effect of bug status: buggy files exhibited a 17\% higher likelihood of receiving comments (coefficient = 0.156, odds ratio = 1.17) and a 6.2\% increase in comment volume when reviewed (coefficient = 0.061, incidence rate ratio = 1.062; both $p < 0.001$). Despite attracting more reviewer attention, these files still led to bug-fixing commits, indicating that reviewer scrutiny, while somewhat aligned with risk signals, is often insufficient to prevent latent defects. Collectively, these findings highlight a review dilution effect in large PRs and expose limitations in the scalability of peer review effectiveness.
\\
\begin{mdframed}[
    linecolor=black!60,
    linewidth=1.5pt,
    backgroundcolor=yellow!8,
    innertopmargin=6pt,
    innerbottommargin=6pt
]
\noindent\textbf{\underline{RQ3 Findings:}} Reviewer comment activity is modestly higher for files that later become latent bugs, but this observable attention does not correspond to reduced latent bug involvement. PR size further dilutes review attention, with each additional modified file reducing the odds that an individual file receives any review comment by approximately 8.7\%. These findings reveal an attention-effectiveness gap, where comment activity is an incomplete proxy for defect-prevention effectiveness, especially in larger PRs.
\end{mdframed}

\section{Discussion}

\subsection{Visible Attention Does Not Guarantee Review Effectiveness}\label{paradox} 
Our investigation of $\mathbf{RQ_3}$ reveals an important gap. While files in higher positions within PRs receive more review comments, an observation consistent with prior studies \cite{fregnan2022first, bagirov2023assessing}, indicating positional bias and cognitive science literature on serial position effects and attention decrement \cite{gershberg1994serial, raaijmakers1981search, wiswede2007serial}, these same files are also disproportionately associated with latent bugs. Over 57\% of such files are later implicated in bug-fixing commits, suggesting that increased commenting does not necessarily translate to effective defect detection. This gap highlights a critical distinction between \textit{review activity} and \textit{review effectiveness}. The gap does not imply that reviewer comments are unhelpful. Rather, it suggests that comments may reflect several forms of engagement, including clarification, maintainability concerns, style issues, formatting problems, or discussion of complex code, not only successful defect detection. A file may attract comments because reviewers recognize that it is important or difficult, but the presence of comments does not guarantee that all relevant defects are identified before integration. This distinction is important because many empirical studies and development teams treat review comments as an observable signal of review effort or coverage~\cite{mcintosh2016empirical,fregnan2022first,bagirov2023assessing}. Our findings suggest that such signals should be interpreted cautiously, especially when the goal is to reason about defect-prevention effectiveness. 

PR size further intensifies this gap through review dilution. The hurdle model shows that each additional modified file reduces the odds that an individual file receives any review comment by approximately 8.7\%. This pattern is consistent with the view that reviewer attention is finite and becomes increasingly difficult to distribute across all files as review scope grows. Cognitive constraints such as working-memory limitations and attention decrement provide a plausible explanation for why reviewers may struggle to maintain thorough inspection across long or scattered file lists~\cite{cowan2001magical,baddeley2009working}. The motivating survey also supports this interpretation, as developers reported that alphabetical ordering can increase context switching, separate related files, and reduce confidence in review completeness~\cite{rahman2026icse}. Therefore, the attention-effectiveness gap should be understood as a socio-technical phenomenon: review outcomes are shaped not only by reviewer effort, but also by PR scope and by how tools structure the inspection sequence.

\subsection{PR Size as a Context-Cognition Trade-off} \label{sec:pr-size-discussion} 

The relationship between PR size and latent bug likelihood reported in $\mathbf{RQ_2}$ introduces nuance beyond the common expectation that smaller PRs are generally easier to review. Prior review guidelines and empirical evidence suggest that compact changes are easier to inspect, while larger reviews can overwhelm reviewers and reduce defect discovery~\cite{cohen2006best,rigby2013convergent}. Our results support the concern that large PRs are vulnerable to attention dilution, but they also show that very small PRs are not necessarily the safest category in terms of latent bug likelihood. Specifically, very small PRs with 2--3 files exhibit elevated latent bug rates, mid-sized PRs around ten files show the lowest observed risk, and large PRs with 20+ files return to elevated rates. This pattern suggests a context-cognition trade-off. Very small PRs may reduce immediate inspection burden, but they can also provide insufficient context for reviewers to evaluate semantic correctness, cross-file dependencies, or the broader impact of a change. At the other extreme, large PRs may provide more context but impose higher cognitive load, increase context switching, and dilute attention across files. Mid-sized PRs may offer a more effective balance by preserving enough contextual information while remaining cognitively manageable. 

This interpretation is consistent with the motivating study \cite{rahman2026icse}. Developers reported that multi-file reviews become difficult when related files are separated, when the logical flow of a change is hard to reconstruct, and when reviewers must repeatedly switch between semantically disconnected files~\cite{rahman2026icse}. These perceptions help explain why PR size should not be treated only as a numeric measure of change volume. Reviewability depends not merely on the number of files, but also on whether the changed files form a coherent unit of reasoning. A small PR that removes too much context may be locally easy to inspect but semantically difficult to validate. A large PR may contain enough context but exceed the reviewer’s capacity to inspect every file with equal care. However, this interpretation does not imply that teams should target a universal ideal PR size, such as exactly ten files. PR reviewability depends on change type~\cite{tao2015partitioning, mcintosh2016empirical}, file relationships~\cite{gonccalves2022explicit, bouraffa2025not}, reviewer expertise~\cite{thongtanunam2015investigating, kononenko2015investigating}, project architecture~\cite{rigby2013convergent}, and the semantic coherence of the change~\cite{baum2017optimal, gonccalves2022explicit}. Rather, our findings suggest that PR size should be treated as an early signal of review vulnerability. For authors, this means that decomposition should preserve meaningful context rather than merely minimizing file count. For reviewers, PR size can help calibrate inspection strategy, prompting more deliberate attention allocation when a PR is either too fragmented or too large. For tool designers, size-based warnings should move beyond generic ``large PR'' alerts and instead help developers identify whether a PR is semantically coherent and reviewable.

\subsection{Triangulating Developer Perceptions with Repository Evidence}
\label{sec:triangulation}

A central value of this article is that it connects practitioner perceptions with repository-level outcome evidence. The motivating study~\cite{rahman2026icse} showed that developers perceive alphabetical ordering as cognitively misaligned: only 10.2\% of respondents considered alphabetical ordering optimal, more than half reported increased context switching and fatigue, and 63.9\% expressed concern or uncertainty that default ordering may cause them to miss bugs. Those findings established the practitioner-facing problem, but they could not determine whether these concerns appear in project history. The present repository-mining study provides the outcome-oriented counterpart. We find that file position is associated with latent bug likelihood ($\mathbf{RQ_1}$), that PR size modulates review vulnerability in a non-linear way ($\mathbf{RQ_2}$), and that review comments do not reliably indicate defect-prevention effectiveness ($\mathbf{RQ_3}$). These results do not prove that alphabetical ordering alone causes defects. However, they show that the concerns raised by developers are not merely subjective preferences about interface convenience. They correspond to measurable patterns in review attention, PR scope, and latent bug involvement.

This triangulation strengthens the overall argument by allowing each study to compensate for what the other cannot establish~\cite{runeson2009guidelines, stol2018abc}. The prior survey study~\cite{rahman2026icse} explains why file ordering matters to developers: it affects how they build context, manage attention, and judge review completeness. The repository-mining study shows why it matters for software quality: file position, PR size, and comment distribution are associated with downstream bug-fixing activity. Together, the two studies position file ordering as a socio-technical factor in code review effectiveness, connecting interface design, reviewer cognition, and quality outcomes \cite{bacchelli2013expectations, sadowski2018modern}.

\subsection{Small Effects Can Matter at Review Scale}
\label{sec:small-effects}

The positional effects observed in this study are statistically significant but modest in magnitude. This should not be interpreted as evidence that file ordering is unimportant. In large-scale software maintenance, small but persistent effects can accumulate across thousands of PRs, reviewers, and projects. Unlike reviewer-level decisions, file ordering is a tool-level default that repeatedly shapes how developers encounter changes across every multi-file PR in a project's history. A slight increase in latent bug likelihood for later-positioned files may appear limited at the individual PR level, but when the same structural pattern repeats across large repositories and long development histories, it can contribute to a meaningful number of escaped defects and follow-up fixes~\cite{mcintosh2014impact,beller2014modern}. Prior methodological work in empirical software engineering emphasizes that practical significance should be interpreted in relation to the scale and context in which an effect operates~\cite{kampenes2007systematic,torkar2021method}. Thus, the value of our findings lies not in claiming that file position is the dominant cause of latent defects, but in showing that file presentation order is a measurable and actionable factor in review effectiveness.

The motivating survey further reinforces why modest effects should be taken seriously. Developers already perceive file ordering as affecting fatigue, context switching, and confidence in review completeness~\cite{rahman2026icse}. When such perceived burdens align with measurable repository patterns, even small statistical effects may become practically meaningful. They point to a recurring tool-mediated friction that may accumulate across routine review work~\cite{kononenko2016code,sadowski2018modern, lenarduzzi2020some}.

\subsection{Toward Review Coverage-Aware Tooling}
\label{sec:coverage-aware-tooling}

Our findings point to an important design implication: code review tools should help reviewers understand and manage review coverage across files. Current interfaces often expose comments, approvals, and file lists, but they provide limited support for determining whether attention has been distributed appropriately across the full PR. A PR with many comments may appear thoroughly reviewed even if those comments are concentrated in only a few files~\cite{fregnan2022first,bagirov2023assessing}. Conversely, files with no comments may be interpreted as unproblematic even when they may have been only lightly inspected or skipped~\cite{baum2017optimal,fregnan2022first}.

Review tools could address this limitation by making coverage distribution more visible. For example, per-file review indicators could show which files have been opened, marked as reviewed, commented on, or left untouched. Attention heatmaps could highlight whether comments and inspection effort are concentrated in early files or in a small subset of the PR. Review progress indicators could help reviewers distinguish between files that were actively inspected and files that remain pending~\cite{baum2017optimal,olewicki2024empirical}. Such mechanisms would make review coverage more explicit and reduce reliance on comment volume as an implicit signal of assurance. They could also warn reviewers when attention is highly concentrated, helping teams detect coverage imbalance before approval.

Our results also support more context-aware file ordering. Instead of relying only on alphabetical ordering, review platforms could support dependency-aware grouping, module-aware grouping, and test-production pairing~\cite{gonccalves2022explicit,bouraffa2025not}, or risk-aware prioritization based on change size, change impact, or historical defect-proneness~\cite{bagirov2023assessing,goccmen2025enhanced}. These approaches should not necessarily replace reviewer judgment. Rather, they can provide alternative views that help reviewers inspect changes in an order that better matches the logical structure of the PR. Importantly, any automated ordering should be explainable and customizable, so that reviewers understand why certain files are prioritized and can adapt the ordering to their task~\cite{olewicki2024empirical,bouraffa2025not}. This is particularly important because reviewer navigation preferences can vary across experience levels, familiarity, and project contexts~\cite{olewicki2024empirical,bouraffa2025not}.

These implications are directly supported by the motivating study~\cite{rahman2026icse}, where developers expressed strong demand for dependency-aware grouping and customizable file ordering. The present findings give those requests additional weight. They suggest that smarter ordering is not merely a usability enhancement, but a potential mechanism for improving attention allocation and review coverage~\cite{fregnan2022first,mcintosh2016empirical}. A useful review interface should therefore support both navigation and assurance: it should help reviewers decide where to begin, what files belong together, which files remain under-reviewed, and whether the review has covered the change in a balanced way.

\subsection{Developer Satisfaction and Trust in Review Outcomes}
\label{sec:satisfaction-trust}

The attention-effectiveness gap also has implications for developer satisfaction and trust in code review. Code review is not only a defect-detection mechanism, but it is also a collaborative process through which authors receive feedback, reviewers invest effort, and teams build shared confidence in code changes~\cite{bacchelli2013expectations,bosu2015characteristics}. When visible review activity does not reliably correspond to defect-prevention effectiveness, both authors and reviewers may experience a mismatch between effort and outcome.

For reviewers, this mismatch may create frustration if careful commenting does not appear to prevent later defects or rework. For authors, dense feedback may create an impression that the change has been comprehensively evaluated, even though some files may have received limited scrutiny. Over time, repeated experiences of missed defects after apparently active reviews could reduce confidence in the review process~\cite{bacchelli2013expectations, 
khatoonabadi2023wasted}. This concern is consistent with prior work showing that developers associate review quality with feedback thoroughness, reviewer familiarity, and the ability to manage context and priorities during review~\cite{kononenko2016code}. Our study does not directly measure satisfaction outcomes in the repository-mining data. However, our prior study shows that developers already perceive alphabetical ordering as contributing to context switching, fatigue, and concern about missed bugs~\cite{rahman2026icse}. The present results provide outcome-level evidence that these concerns are plausible and worthy of tool-design attention.

Accordingly, improving file ordering and review coverage awareness is not only a technical quality concern. It is also a way to support a more trustworthy and satisfying review process. Interfaces that help reviewers allocate attention deliberately, preserve context, and make coverage visible~\cite{baum2017optimal, olewicki2024empirical} can reduce uncertainty for both reviewers and authors. Such support may strengthen the perceived usefulness of review feedback and improve confidence that multi-file PRs have been evaluated thoroughly~\cite{bosu2015characteristics,kononenko2016code}.

\noindent Finally, our findings should be interpreted as evidence of association rather than causation. We do not claim that alphabetical ordering alone causes latent bugs or that reviewer comments are ineffective. Rather, our results show that file presentation order, PR size, and review attention are empirically associated with review-effectiveness signals at scale. Combined with practitioner evidence from the motivating study~\cite{rahman2026icse}, these findings suggest that alphabetical file ordering should not be treated as a neutral interface default, but as a structural feature of review tooling that can shape attention allocation and review coverage.


\section{Research Implications}

\subsection{Implications for Practitioners and Tool Builders}

Our findings yield three actionable implications for development teams, maintainers, and review-tool designers. First, review workflows should include simple checks to ensure that all files in a PR receive enough reviewer attention. Because reviewer attention is partly shaped by file ordering and PR scope~\cite{fregnan2022first, bagirov2023assessing, rahman2026icse}, comment volume alone may provide an incomplete, and sometimes misleading, picture of review coverage. Lightweight mechanisms such as per-file checklists, structured ``reviewed'' indicators, or explicit file-level acknowledgements can help ensure that each file receives deliberate consideration rather than incidental scrutiny~\cite{baum2017optimal, olewicki2024empirical}. Second, PR-size guidance should be treated as a quality practice rather than merely a convenience norm. The observed non-linear risk pattern suggests that simply minimizing PR size is insufficient. Teams should encourage context-sensitive decomposition: PRs should be small enough to remain cognitively manageable, but large enough to preserve the contextual information reviewers need to assess semantic correctness and cross-file dependencies~\cite{cohen2006best, rigby2013convergent, gonccalves2022explicit}. Such guidance can be operationalized through contribution guidelines, PR templates, and reviewer checklists that frame reviewability expectations as a shared quality responsibility. Third, review platforms should make coverage distribution more transparent. Since sparse commenting on a file can reflect correctness, silent inspection, or omission, comment absence is an ambiguous signal that teams and tools should not treat as implicit confirmation of thoroughness~\cite{bacchelli2013expectations, bosu2015characteristics}. Platform support could include per-file engagement summaries, indicators for files opened or marked as reviewed, and warnings when comments are concentrated in only a small subset of the PR~\cite{mcintosh2016empirical, olewicki2024empirical}. Such features would reduce false assurance, support mutual accountability between authors and reviewers, and help reviewers allocate attention more deliberately across multi-file PRs.

\subsection{Implications for Researchers}

This study also motivates several research directions. First, the attention--effectiveness gap warrants deeper investigation into the nature of review attention itself. Future work should distinguish whether observed comments primarily address surface-level concerns such as style and formatting, or reflect deeper semantic evaluation. Qualitative coding of review comments~\cite{mantyla2008types}, controlled experiments~\cite{fregnan2022first}, and eye-tracking 
studies~\cite{huang2020biases} could help explain when and why visible review activity translates into effective defect detection and when it does not. Second, the baselines established in this study enable principled evaluation of alternative file-ordering strategies. Subsequent studies can compare dependency-aware, risk-prioritized, module-based, or semantically grouped orderings against alphabetical ordering to assess their effects on coverage uniformity, reviewer effort distribution, cognitive load, and latent defect outcomes. Such studies should examine not only whether alternative orderings increase reviewer attention, but also whether they improve defect-prevention effectiveness. Our dataset and replication package provide the necessary infrastructure for such comparisons. Third, effective PR decomposition remains an open problem. Our findings show that neither minimizing nor maximizing file count reliably reduces latent bug risk. Future work should investigate what decomposition criteria, such as semantic coherence, dependency containment, or change-type homogeneity, best preserve the contextual information reviewers need while keeping PRs within a cognitively manageable scope~\cite{tao2015partitioning, rigby2013convergent, gonccalves2022explicit}. This could move PR-size guidance beyond simple file-count thresholds toward evidence-based models of reviewability. Fourth, future research should also examine the developer-experience consequences of structural coverage imbalance directly. Survey, interview, and field-intervention studies can investigate whether developers perceive positional coverage gaps, how awareness of such gaps affects review behavior, and whether transparency interventions such as per-file summaries, review progress indicators, or attention heatmaps improve reviewer confidence, author trust, and perceived review fairness. This direction 
matters because review effectiveness is not only a technical outcome, but also a socio-technical experience shaped by confidence, trust, and perceived thoroughness.

\section{Threats to Validity}

\subsection{Construct Validity}

Our identification of bug-fixing PRs relies on a keyword-based heuristic using terms such as `bug' and `fix' in commit messages, following Mockus et al.~\cite{mockus2000identifying}. While this approach is widely adopted and validated in prior studies~\cite{barbour2013empirical, islam2017comparative}, it may miss bug-fixing commits with unconventional wording and may include 
commits where the keyword does not correspond to an actual defect fix. To reduce such false positives, we excluded commits whose messages combined bug keywords with refactoring or documentation indicators, following Islam~et~al.~\cite{islam2017comparative}. To assess precision, we manually validated 384 randomly sampled bug-fixing and non-bug-fixing commit messages, corresponding to a 95\% confidence level and 5\% margin of error. Two evaluators with more than 12 years of software development experience achieved strong inter-rater agreement (Cohen's $\kappa = 0.97$), with 96.61\% and 99.18\% alignment with our labels, respectively. We also define \textit{latent bug files} using a historical proxy: a file is labeled latent buggy when it appears in a bug-fixing PR immediately after a prior modification. This assumption builds on prior research indicating that bug-fixing commits typically address one or more defects in the software~\cite{tufano2019empirical, vieira2019reports, herbold2022fine}. However, this proxy does not provide ground truth defect attribution. A subsequent bug-fixing change may address a defect introduced before the earlier PR, may be related to later requirements, or may reflect maintenance activity rather than a review-missed defect. Therefore, we interpret latent bug labels as a practical proxy for post-review defect involvement rather than as direct evidence of defect introduction.

A further construct threat concerns our operationalization of file position. We approximate file position using the default path-based presentation order available from repository and PR metadata. This reflects the ordering used by common review tools, but it does not necessarily capture the exact order in which every reviewer inspected files. Reviewers can manually jump between files, collapse or expand diffs, filter viewed files, or rely on local tools outside the platform interface. Our motivating study partly contextualizes this limitation: 42.4\% of reviewers reported following the default alphabetical order, while 57.6\% reported using alternative, context-driven navigation strategies~\cite{rahman2026icse}. Thus, file position in this study should be interpreted as the default presentation position and a meaningful interface-level exposure, not as a precise trace of each reviewer’s cognitive inspection sequence. Finally, we operationalize review activity using the number of inline comments per file. This measure captures observable interaction but not the depth or quality of cognitive engagement. High comment volume can reflect low-cost feedback, such as style, naming, formatting, or clarification remarks, and may not correspond to semantic evaluation or defect discovery. Conversely, a file with no comments may still have been carefully inspected. Therefore, our RQ3 findings should be interpreted as evidence about visible review interaction rather than a direct measurement of reviewer cognition or semantic scrutiny.

\subsection{Internal Validity}

The observed association between file position and latent bug likelihood may partially reflect file-intrinsic properties that correlate with alphabetical naming conventions rather than a positional mechanism per se. For instance, files that appear early under alphabetical ordering may more often be core abstractions or shared utilities, which are typically more change-prone and highly coupled, and therefore more defect-prone, independent of review ordering. We partially mitigate this risk in two ways. First, we model PR size jointly with bug status in our regression analyses, reducing confounding from scope-related review dynamics. Second, we repeat the analyses on a restricted subset of programming and logic implementation files (filtered by extension) and observe consistent patterns, suggesting the results are not driven solely by heterogeneous file types. However, we do not explicitly control for file-level complexity and process covariates (e.g., churn, coupling, complexity), which are known to predict faults independently of review and process factors \cite{di2017developer, nagappan2005use}. Future work should incorporate these covariates to isolate positional effects from baseline defect proneness better.

Our latent bug identification also relies on a sequential assumption: a file is marked latent buggy when it reappears in a bug-fixing context immediately after an earlier PR. This assumption may be violated when multiple PRs are developed in parallel, when branches diverge, or when the subsequent bug fix addresses an older 
defect unrelated to the immediately preceding PR. We mitigate this risk by applying the labeling procedure uniformly over commit-level file histories within each repository, consistent with established sequential SZZ-inspired approaches~\cite{da2016framework, wen2019exploring}. Nevertheless, the resulting labels should be interpreted as probabilistic indicators of post-review defect involvement rather than causal links between a specific review and a specific defect. Furthermore, our study is scoped specifically to PR-based review workflows, where file ordering and review activity are observable. Direct commits outside PRs are not associated with the same review interface and therefore do not expose file-ordering or comment-based review signals. Many mature projects use branch protection or contribution policies\footnote{\url{https://docs.github.com/en/repositories/configuring-branches-and-merges-in-your-repository/managing-protected-branches/managing-a-branch-protection-rule}}\footnote{\url{https://cwiki.apache.org/confluence/display/solr/CommitPolicy}}\footnote{\url{https://devguide.python.org/core-team/committing/}}\footnote{\url{https://www.kubernetes.dev/docs/guide/contributing}} that encourage substantive changes to flow through reviewed PRs. Nevertheless, intervening non-PR commits may still occur and could affect the attribution of latent bug involvement.

\subsection{External Validity}

Our external validity is bounded by context, ecosystem composition, and the ordering condition studied. The dataset consists of 182 open-source GitHub projects, including Apache and popular non-Apache repositories. These projects use asynchronous and often volunteer-driven review practices, which may differ from industrial review in reviewer assignment, turnaround expectations, ownership structures, and tool support. Prior practitioner evidence suggests that industrial workflows often involve more structured reviewer assignment and smaller, more targeted changes~\cite{rigby2013convergent, bacchelli2013expectations}, so replication in proprietary or enterprise settings remains necessary. Moreover, our analysis includes only merged PRs. This choice is appropriate for studying post-review defect involvement in integrated changes, but it excludes rejected, abandoned, or still-open PRs. Such PRs may contain different review dynamics, defect patterns, or comment distributions. Therefore, our findings generalize primarily to code changes that pass review and enter the project history. 

Although the dataset spans five programming languages, it is weighted toward Java and toward projects with established GitHub-based workflows. Java's package-based directory conventions interact with alphabetical path ordering differently than flatter ecosystems such as Python, which may affect how positional effects manifest across languages. Additionally, GitHub projects vary substantially in community norms, review culture, and activity levels across domains, which may limit generalizability beyond the open-source context studied~\cite{kalliamvakou2014promises}. Furthermore, our inclusion criteria favor mature, active, and popular repositories, systematically excluding projects with weaker review cultures or less experienced contributors that may exhibit stronger positional effects. Our findings therefore likely represent a conservative estimate of positional bias, and replication across projects of varying maturity and review quality remains necessary~\cite{kalliamvakou2014promises, cosentino2017systematic}.

\section{Conclusion}

This study examined whether the default alphabetical file order in contemporary code review tools is associated with review effectiveness at scale. Building on our motivating prior study, which showed that developers perceive alphabetical ordering as cognitively misaligned with multi-file PR review, we investigated whether these concerns correspond to repository-level outcomes. Using 330,343 multi-file pull requests comprising 756,814 file instances from 182 GitHub projects, we find statistically significant, although modest, associations between file position, PR size, review activity, and latent bug likelihood. Later-positioned files show slightly higher latent bug rates, increasing from 56.7\% at position 1 to 61.5\% at position 30, while mid-sized PRs around ten files show the lowest observed risk. Review attention also decreases as PR size grows, with each additional modified file reducing the odds that a file receives any review comment by approximately 8.7\%. Together, these findings expose an attention-effectiveness gap: files receiving more comments are not correspondingly less likely to appear in later bug-fixing changes, indicating that observable review activity is not equivalent to defect prevention. Combined with our prior survey-based study, our results suggest that alphabetical ordering should not be treated as a neutral interface default because it can shape attention allocation, review coverage, and confidence in review outcomes. These findings motivate context-aware ordering, dependency-aware grouping, risk-aware prioritization, and per-file coverage indicators, while encouraging teams to treat PR size and review coverage as shared quality concerns rather than relying on comment volume alone. More broadly, this study positions file ordering as a socio-technical factor in code review effectiveness, connecting tool design, reviewer cognition, developer experience, and downstream software quality.

\section*{Acknowledgements}
This research is supported in part by the Natural Sciences and Engineering Research Council of Canada (NSERC) Discovery Grants program and by the industry-stream NSERC CREATE in Software Analytics Research (SOAR).

\bibliographystyle{ACM-Reference-Format}
\bibliography{sample-base}

@inproceedings{tao2015partitioning,
  title={Partitioning composite code changes to facilitate code review},
  author={Tao, Yida and Kim, Sunghun},
  booktitle={2015 IEEE/ACM 12th Working Conference on Mining Software Repositories},
  pages={180--190},
  year={2015},
  organization={IEEE}
}

@inproceedings{bavota2015four,
  title={Four eyes are better than two: On the impact of code reviews on software quality},
  author={Bavota, Gabriele and Russo, Barbara},
  booktitle={2015 IEEE International Conference on Software Maintenance and Evolution (ICSME)},
  pages={81--90},
  year={2015},
  organization={IEEE}
}

@inproceedings{barnett2015helping,
  title={Helping developers help themselves: Automatic decomposition of code review changesets},
  author={Barnett, Mike and Bird, Christian and Brunet, Jo{\~a}o and Lahiri, Shuvendu K},
  booktitle={2015 IEEE/ACM 37th IEEE International Conference on Software Engineering (ICSE)},
  volume={1},
  pages={134--144},
  year={2015},
  organization={IEEE}
}

@article{wang2015comparative,
  title={Comparative case studies of open source software peer review practices},
  author={Wang, Jing and Shih, Patrick C and Wu, Yu and Carroll, John M},
  journal={Information and Software Technology},
  volume={67},
  pages={1--12},
  year={2015},
  publisher={Elsevier}
}

@inproceedings{bosu2013impact,
  title={Impact of peer code review on peer impression formation: A survey},
  author={Bosu, Amiangshu and Carver, Jeffrey C},
  booktitle={2013 ACM/IEEE International Symposium on Empirical Software Engineering and Measurement (ESEM)},
  pages={133--142},
  year={2013},
  organization={IEEE}
}

@article{bosu2016process,
  title={Process aspects and social dynamics of contemporary code review: Insights from open source development and industrial practice at microsoft},
  author={Bosu, Amiangshu and Carver, Jeffrey C and Bird, Christian and Orbeck, Jonathan and Chockley, Christopher},
  journal={IEEE Transactions on Software Engineering},
  volume={43},
  number={1},
  pages={56--75},
  year={2016},
  publisher={IEEE}
}

@inproceedings{morales2015code,
  title={Do code review practices impact design quality? a case study of the qt, vtk, and itk projects},
  author={Morales, Rodrigo and McIntosh, Shane and Khomh, Foutse},
  booktitle={2015 IEEE 22nd international conference on software analysis, evolution, and reengineering (SANER)},
  pages={171--180},
  year={2015},
  organization={IEEE}
}

@article{mcintosh2016empirical,
  title={An empirical study of the impact of modern code review practices on software quality},
  author={McIntosh, Shane and Kamei, Yasutaka and Adams, Bram and Hassan, Ahmed E},
  journal={Empirical Software Engineering},
  volume={21},
  number={5},
  pages={2146--2189},
  year={2016},
  publisher={Springer}
}

@inproceedings{bacchelli2013expectations,
  title={Expectations, outcomes, and challenges of modern code review},
  author={Bacchelli, Alberto and Bird, Christian},
  booktitle={2013 35th International Conference on Software Engineering (ICSE)},
  pages={712--721},
  year={2013},
  organization={IEEE}
}

@inproceedings{sadowski2018modern,
  title={Modern code review: a case study at google},
  author={Sadowski, Caitlin and S{\"o}derberg, Emma and Church, Luke and Sipko, Michal and Bacchelli, Alberto},
  booktitle={Proceedings of the 40th international conference on software engineering: Software engineering in practice},
  pages={181--190},
  year={2018}
}

@book{cohen2006best,
  title={Best kept secrets of peer code review},
  author={Cohen, Jason and Teleki, Steven and Brown, Eric},
  year={2006},
  publisher={Smart Bear Incorporated}
}

@inproceedings{mukadam2013gerrit,
  title={Gerrit software code review data from android},
  author={Mukadam, Murtuza and Bird, Christian and Rigby, Peter C},
  booktitle={2013 10th Working Conference on Mining Software Repositories (MSR)},
  pages={45--48},
  year={2013},
  organization={IEEE}
}

@book{loeliger2012version,
  title={Version Control with Git: Powerful tools and techniques for collaborative software development},
  author={Loeliger, Jon and McCullough, Matthew},
  year={2012},
  publisher={" O'Reilly Media, Inc."}
}

@incollection{fagan2002design,
  title={Design and code inspections to reduce errors in program development},
  author={Fagan, Michael},
  booktitle={Software pioneers},
  pages={575--607},
  year={2002},
  publisher={Springer}
}

@inproceedings{bosu2012peer,
  title={Peer code review in open source communities using reviewboard},
  author={Bosu, Amiangshu and Carver, Jeffrey C},
  booktitle={Proceedings of the ACM 4th annual workshop on Evaluation and usability of programming languages and tools},
  pages={17--24},
  year={2012}
}

@article{tsotsis2011meet,
  title={Meet phabricator, the witty code review tool built inside facebook},
  author={Tsotsis, Alexia},
  journal={Retrieved November},
  volume={30},
  pages={2021},
  year={2011}
}

@article{blischak2016quick,
  title={A quick introduction to version control with Git and GitHub},
  author={Blischak, John D and Davenport, Emily R and Wilson, Greg},
  journal={PLoS computational biology},
  volume={12},
  number={1},
  pages={1-18},
  year={2016},
  publisher={Public Library of Science}
}

@article{davila2021systematic,
  title={A systematic literature review and taxonomy of modern code review},
  author={Davila, Nicole and Nunes, Ingrid},
  journal={Journal of Systems and Software},
  volume={177},
  pages={110951},
  year={2021},
  publisher={Elsevier}
}

@inproceedings{baum2017optimal,
  title={On the optimal order of reading source code changes for review},
  author={Baum, Tobias and Schneider, Kurt and Bacchelli, Alberto},
  booktitle={2017 IEEE international conference on software maintenance and evolution (ICSME)},
  pages={329--340},
  year={2017},
  organization={IEEE}
}

@inproceedings{fregnan2022first,
  title={First come first served: The impact of file position on code review},
  author={Fregnan, Enrico and Braz, Larissa and D'Ambros, Marco and {\c{C}}al{\i}kl{\i}, G{\"u}l and Bacchelli, Alberto},
  booktitle={Proceedings of the 30th ACM joint european software engineering conference and symposium on the foundations of software engineering},
  pages={483--494},
  year={2022}
}

@article{olewicki2024empirical,
  title={An empirical study on code review activity prediction and its impact in practice},
  author={Olewicki, Doriane and Habchi, Sarra and Adams, Bram},
  journal={Proceedings of the ACM on Software Engineering},
  volume={1},
  number={FSE},
  pages={2238--2260},
  year={2024},
  publisher={ACM New York, NY, USA}
}

@article{bouraffa2025not,
  title={Not One to Rule Them All: Mining Meaningful Code Review Orders From GitHub},
  author={Bouraffa, Abir and Brandt, Carolin and Zaidmann, Andy and Maalej, Walid},
  journal={In Proceedings of the 29th International Conference on Evaluation and Assessment in Software Engineering (EASE 2025), Istanbul, Türkiye},
  year={2025},
  organization={ACM}
}

@inproceedings{bagirov2023assessing,
  title={Assessing the impact of file ordering strategies on code review process},
  author={Bagirov, Farid and Derakhshanfar, Pouria and Kalina, Alexey and Kartysheva, Elena and Kovalenko, Vladimir},
  booktitle={Proceedings of the 27th International Conference on Evaluation and Assessment in Software Engineering (EASE)},
  pages={188--191},
  year={2023}
}

@article{rahman2026icse,
  title={Breaking the Alphabet: Rethinking File Ordering in Code Review},
  author={Rahman, Md Shamimur and Codabux, Zadia and Roy, Chanchal K},
  journal={Proceedings of the 48th International Conference on Software Engineering (ICSE)},
  year={2026}
}

@article{baum2019cognitive,
  title={Cognitive-support code review tools: improved efficiency of change-based code review by guiding and assisting reviewers},
  author={Baum, Tobias},
  year={2019},
  publisher={Hannover: Institutionelles Repositorium der Universit{\"a}t Hannover}
}

@article{mohanani2018cognitive,
  title={Cognitive biases in software engineering: A systematic mapping study},
  author={Mohanani, Rahul and Salman, Iflaah and Turhan, Burak and Rodr{\'\i}guez, Pilar and Ralph, Paul},
  journal={IEEE Transactions on Software Engineering},
  volume={46},
  number={12},
  pages={1318--1339},
  year={2018},
  publisher={IEEE}
}

@inproceedings{spadini2020primers,
  title={Primers or reminders? The effects of existing review comments on code review},
  author={Spadini, Davide and {\c{C}}alikli, G{\"u}l and Bacchelli, Alberto},
  booktitle={Proceedings of the ACM/IEEE 42nd International Conference on Software Engineering},
  pages={1171--1182},
  year={2020}
}

@inproceedings{mockus2000identifying,
  title={Identifying reasons for software changes using historic databases},
  author={Mockus and Votta},
  booktitle={Proceedings 2000 international conference on software maintenance},
  pages={120--130},
  year={2000},
  organization={IEEE}
}

@inproceedings{islam2017comparative,
  title={A Comparative Study of Software Bugs in Clone and Non-Clone Code.},
  author={Islam, Judith F and Mondal, Manishankar and Roy, Chanchal K and Schneider, Kevin A},
  booktitle={SEKE},
  pages={436--443},
  year={2017}
}

@inproceedings{rigby2013convergent,
  title={Convergent contemporary software peer review practices},
  author={Rigby, Peter C and Bird, Christian},
  booktitle={Proceedings of the 2013 9th joint meeting on foundations of software engineering},
  pages={202--212},
  year={2013}
}

@article{baum2019associating,
  title={Associating working memory capacity and code change ordering with code review performance},
  author={Baum, Tobias and Schneider, Kurt and Bacchelli, Alberto},
  journal={Empirical Software Engineering},
  volume={24},
  pages={1762--1798},
  year={2019},
  publisher={Springer}
}

@inproceedings{gonccalves2020explicit,
  title={Do explicit review strategies improve code review performance?},
  author={Gon{\c{c}}alves, Pavl{\'\i}na Wurzel and Fregnan, Enrico and Baum, Tobias and Schneider, Kurt and Bacchelli, Alberto},
  booktitle={Proceedings of the 17th international conference on mining software repositories},
  pages={606--610},
  year={2020}
}

@article{gonccalves2022explicit,
  title={Do explicit review strategies improve code review performance? Towards understanding the role of cognitive load},
  author={Gon{\c{c}}alves, Pavl{\'\i}na Wurzel and Fregnan, Enrico and Baum, Tobias and Schneider, Kurt and Bacchelli, Alberto},
  journal={Empirical Software Engineering (EMSE)},
  volume={27},
  number={4},
  pages={99},
  year={2022},
  publisher={Springer}
}

@inproceedings{chattopadhyay2020tale,
  title={A tale from the trenches: cognitive biases and software development},
  author={Chattopadhyay, Souti and Nelson, Nicholas and Au, Audrey and Morales, Natalia and Sanchez, Christopher and Pandita, Rahul and Sarma, Anita},
  booktitle={Proceedings of the ACM/IEEE 42nd International Conference on Software Engineering},
  pages={654--665},
  year={2020}
}

@inproceedings{huang2020biases,
  title={Biases and differences in code review using medical imaging and eye-tracking: genders, humans, and machines},
  author={Huang, Yu and Leach, Kevin and Sharafi, Zohreh and McKay, Nicholas and Santander, Tyler and Weimer, Westley},
  booktitle={Proceedings of the 28th ACM joint meeting on European software engineering conference and symposium on the foundations of software engineering},
  pages={456--468},
  year={2020}
}

@article{thongtanunam2020review,
  title={Review dynamics and their impact on software quality},
  author={Thongtanunam, Patanamon and Hassan, Ahmed E},
  journal={IEEE Transactions on Software Engineering},
  volume={47},
  number={12},
  pages={2698--2712},
  year={2020},
  publisher={IEEE}
}

@inproceedings{jetzen2025towards,
  title={Towards debiasing code review support},
  author={Jetzen, Tobias and Devroey, Xavier and Matton, Nicolas and Vanderose, Beno{\^\i}t},
  booktitle={2025 IEEE/ACM 18th International Conference on Cooperative and Human Aspects of Software Engineering (CHASE)},
  pages={143--148},
  year={2025},
  organization={IEEE}
}

@article{murdock1962serial,
  title={The serial position effect of free recall.},
  author={Murdock Jr, Bennet B},
  journal={Journal of experimental psychology},
  volume={64},
  number={5},
  pages={482},
  year={1962},
  publisher={American Psychological Association}
}

@book{vandenbos2007apa,
  title={APA dictionary of psychology.},
  author={VandenBos, Gary R},
  year={2007},
  publisher={American Psychological Association}
}

@article{hendrick1973attention,
  title={Attention decrement, temporal variation, and the primacy effect in impression formation},
  author={Hendrick, Clyde and Costantini, Arthur F and McGarry, James and McBride, Keith},
  journal={Memory \& cognition},
  volume={1},
  pages={193--195},
  year={1973},
  publisher={Springer}
}

@article{baddeley2003working,
  title={Working memory: looking back and looking forward},
  author={Baddeley, Alan},
  journal={Nature reviews neuroscience},
  volume={4},
  number={10},
  pages={829--839},
  year={2003},
  publisher={Nature Publishing Group UK London}
}

@article{wilhelm2013working,
  title={What is working memory capacity, and how can we measure it?},
  author={Wilhelm, Oliver and Hildebrandt, Andrea and Oberauer, Klaus},
  journal={Frontiers in psychology},
  volume={4},
  pages={433},
  year={2013},
  publisher={Frontiers Media SA}
}

@article{cowan2001magical,
  title={The magical number 4 in short-term memory: A reconsideration of mental storage capacity},
  author={Cowan, Nelson},
  journal={Behavioral and brain sciences},
  volume={24},
  number={1},
  pages={87--114},
  year={2001},
  publisher={Cambridge University Press}
}

@article{cowan2010magical,
  title={The magical mystery four: How is working memory capacity limited, and why?},
  author={Cowan, Nelson},
  journal={Current directions in psychological science},
  volume={19},
  number={1},
  pages={51--57},
  year={2010},
  publisher={Sage Publications Sage CA: Los Angeles, CA}
}

@article{bergersen2011programming,
  title={Programming skill, knowledge, and working memory among professional software developers from an investment theory perspective},
  author={Bergersen, Gunnar Rye and Gustafsson, Jan-Eric},
  journal={Journal of individual Differences},
  year={2011},
  publisher={Hogrefe Publishing}
}

@inproceedings{abid2019developer,
  title={Developer reading behavior while summarizing java methods: Size and context matters},
  author={Abid, Nahla J and Sharif, Bonita and Dragan, Natalia and Alrasheed, Hend and Maletic, Jonathan I},
  booktitle={2019 IEEE/ACM 41st International Conference on Software Engineering (ICSE)},
  pages={384--395},
  year={2019},
  organization={IEEE}
}

@article{rodeghero2015eye,
  title={An eye-tracking study of java programmers and application to source code summarization},
  author={Rodeghero, Paige and Liu, Cheng and McBurney, Paul W and McMillan, Collin},
  journal={IEEE Transactions on Software Engineering},
  volume={41},
  number={11},
  pages={1038--1054},
  year={2015},
  publisher={IEEE}
}

@inproceedings{al2021novice,
  title={From novice to expert: Analysis of token level effects in a longitudinal eye tracking study},
  author={Al Madi, Naser and Peterson, Cole S and Sharif, Bonita and Maletic, Jonathan I},
  booktitle={2021 IEEE/ACM 29th International Conference on Program Comprehension (ICPC)},
  pages={172--183},
  year={2021},
  organization={IEEE}
}

@inproceedings{siy2001does,
  title={Does the modern code inspection have value?},
  author={Siy, Harvey and Votta, Lawrence},
  booktitle={Proceedings IEEE International Conference on Software Maintenance. ICSM 2001},
  pages={281--289},
  year={2001},
  organization={IEEE}
}

@article{mantyla2008types,
  title={What types of defects are really discovered in code reviews?},
  author={M{\"a}ntyl{\"a}, Mika V and Lassenius, Casper},
  journal={IEEE Transactions on Software Engineering},
  volume={35},
  number={3},
  pages={430--448},
  year={2008},
  publisher={IEEE}
}

@inproceedings{yu2023security,
  title={Security defect detection via code review: A study of the openstack and qt communities},
  author={Yu, Jiaxin and Fu, Liming and Liang, Peng and Tahir, Amjed and Shahin, Mojtaba},
  booktitle={2023 ACM/IEEE International Symposium on Empirical Software Engineering and Measurement (ESEM)},
  pages={1--12},
  year={2023},
  organization={IEEE}
}

@article{tufano2024code,
  title={Code review automation: strengths and weaknesses of the state of the art},
  author={Tufano, Rosalia and Dabi{\'c}, Ozren and Mastropaolo, Antonio and Ciniselli, Matteo and Bavota, Gabriele},
  journal={IEEE Transactions on Software Engineering},
  volume={50},
  number={2},
  pages={338--353},
  year={2024},
  publisher={IEEE}
}

@article{tufano2025deep,
  title={Deep Learning-based Code Reviews: A Paradigm Shift or a Double-Edged Sword?},
  author={Tufano, Rosalia and Martin-Lopez, Alberto and Tayeb, Ahmad and Haiduc, Sonia and Bavota, Gabriele and others},
  journal={IEEE Transactions on Software Engineering},
  pages={1640--1652},
  year={2025},
  publisher={IEEE}
}

@article{goccmen2025enhanced,
  title={Enhanced code reviews using pull request based change impact analysis},
  author={G{\"o}{\c{c}}men, Ismail Sergen and Cezayir, Ahmed Salih and T{\"u}z{\"u}n, Eray},
  journal={Empirical Software Engineering},
  volume={30},
  number={3},
  pages={64},
  year={2025},
  publisher={Springer}
}

@inproceedings{thongtanunam2015investigating,
  title={Investigating code review practices in defective files: An empirical study of the qt system},
  author={Thongtanunam, Patanamon and McIntosh, Shane and Hassan, Ahmed E and Iida, Hajimu},
  booktitle={2015 ieee/acm 12th working conference on mining software repositories},
  pages={168--179},
  year={2015},
  organization={IEEE}
}

@inproceedings{kalliamvakou2014promises,
  title={The promises and perils of mining github},
  author={Kalliamvakou, Eirini and Gousios, Georgios and Blincoe, Kelly and Singer, Leif and German, Daniel M and Damian, Daniela},
  booktitle={Proceedings of the 11th working conference on mining software repositories},
  pages={92--101},
  year={2014}
}

@inproceedings{gousios2014dataset,
  title={A dataset for pull-based development research},
  author={Gousios, Georgios and Zaidman, Andy},
  booktitle={Proceedings of the 11th Working Conference on Mining Software Repositories},
  pages={368--371},
  year={2014}
}

@article{borges2018s,
  title={What’s in a github star? understanding repository starring practices in a social coding platform},
  author={Borges, Hudson and Valente, Marco Tulio},
  journal={Journal of Systems and Software},
  volume={146},
  pages={112--129},
  year={2018},
  publisher={Elsevier}
}

@inproceedings{elazhary2019not,
  title={Do as i do, not as i say: Do contribution guidelines match the github contribution process?},
  author={Elazhary, Omar and Storey, Margaret-Anne and Ernst, Neil and Zaidman, Andy},
  booktitle={2019 IEEE International Conference on Software Maintenance and Evolution (ICSME)},
  pages={286--290},
  year={2019},
  organization={IEEE}
}

@article{barbour2013empirical,
  title={An empirical study of faults in late propagation clone genealogies},
  author={Barbour, Liliane and Khomh, Foutse and Zou, Ying},
  journal={Journal of Software: Evolution and Process},
  volume={25},
  number={11},
  pages={1139--1165},
  year={2013},
  publisher={Wiley Online Library}
}

@article{cohen1960coefficient,
  title={A coefficient of agreement for nominal scales},
  author={Cohen, Jacob},
  journal={Educational and psychological measurement},
  volume={20},
  number={1},
  pages={37--46},
  year={1960},
  publisher={Sage Publications Sage CA: Thousand Oaks, CA}
}

@inproceedings{mcintosh2014impact,
  title={The impact of code review coverage and code review participation on software quality: A case study of the qt, vtk, and itk projects},
  author={McIntosh, Shane and Kamei, Yasutaka and Adams, Bram and Hassan, Ahmed E},
  booktitle={Proceedings of the 11th working conference on mining software repositories},
  pages={192--201},
  year={2014}
}

@book{powers2008statistical,
  title={Statistical methods for categorical data analysis},
  author={Powers, Daniel and Xie, Yu},
  year={2008},
  publisher={Emerald Group Publishing}
}

@article{cressie1986use,
  title={How to use the two sample t-test},
  author={Cressie, NAC and Whitford, HJ},
  journal={Biometrical Journal},
  volume={28},
  number={2},
  pages={131--148},
  year={1986},
  publisher={Wiley Online Library}
}

@article{mcknight2010mannn,
  title={Mann-Whitney U Test},
  author={McKnight, Patrick E and Najab, Julius},
  journal={The Corsini encyclopedia of psychology},
  pages={1--1},
  year={2010},
  publisher={Wiley Online Library}
}

@incollection{benesty2009pearson,
  title={Pearson correlation coefficient},
  author={Benesty, Jacob and Chen, Jingdong and Huang, Yiteng and Cohen, Israel},
  booktitle={Noise reduction in speech processing},
  pages={1--4},
  year={2009},
  publisher={Springer}
}

@article{sedgwick2014spearman,
  title={Spearman’s rank correlation coefficient},
  author={Sedgwick, Philip},
  journal={Bmj},
  volume={349},
  year={2014},
  publisher={British Medical Journal Publishing Group}
}

@article{diener2010cohen,
  title={Cohen's d},
  author={Diener, Marc J},
  journal={The Corsini encyclopedia of psychology},
  pages={1--1},
  year={2010},
  publisher={Wiley Online Library}
}

@article{mchugh2013chi,
  title={The chi-square test of independence},
  author={McHugh, Mary L},
  journal={Biochemia medica},
  volume={23},
  number={2},
  pages={143--149},
  year={2013},
  publisher={Medicinska naklada}
}

@book{cramer1999mathematical,
  title={Mathematical methods of statistics},
  author={Cram{\'e}r, Harald},
  volume={26},
  year={1999},
  publisher={Princeton university press}
}

@incollection{agresti2011categorical,
  title={Categorical data analysis},
  author={Agresti, Alan and Kateri, Maria},
  booktitle={International encyclopedia of statistical science},
  pages={206--208},
  year={2011},
  publisher={Springer}
}

@article{feng2021comparison,
  title={A comparison of zero-inflated and hurdle models for modeling zero-inflated count data},
  author={Feng, Cindy Xin},
  journal={Journal of statistical distributions and applications},
  volume={8},
  number={1},
  pages={8},
  year={2021},
  publisher={Springer}
}

@book{cameron2013regression,
  title={Regression analysis of count data},
  author={Cameron, Adrian Colin and Trivedi, Pravin K},
  number={53},
  year={2013},
  publisher={Cambridge university press}
}

@article{rosenthal1994parametric,
  title={Parametric measures of effect size},
  author={Rosenthal, Robert and Cooper, Harris and Hedges, Larry and others},
  journal={The handbook of research synthesis},
  volume={621},
  number={2},
  pages={231--244},
  year={1994},
  publisher={New York}
}

@article{akoglu2018user,
  title={User's guide to correlation coefficients},
  author={Akoglu, Haldun},
  journal={Turkish journal of emergency medicine},
  volume={18},
  number={3},
  pages={91--93},
  year={2018},
  publisher={Elsevier}
}

@article{sobel1959group,
  title={Group testing to eliminate efficiently all defectives in a binomial sample},
  author={Sobel, Milton and Groll, Phyllis A},
  journal={Bell System Technical Journal},
  volume={38},
  number={5},
  pages={1179--1252},
  year={1959},
  publisher={Wiley Online Library}
}

@article{mckight2010kruskal,
  title={Kruskal-wallis test},
  author={McKight, Patrick E and Najab, Julius},
  journal={The corsini encyclopedia of psychology},
  pages={1--1},
  year={2010},
  publisher={Wiley Online Library}
}

@article{abdi2007kendall,
  title={The Kendall rank correlation coefficient},
  author={Abdi, Herv{\'e}},
  journal={Encyclopedia of Measurement and Statistics. Sage, Thousand Oaks, CA},
  pages={508--510},
  year={2007}
}

@article{gershberg1994serial,
  title={Serial position effects in implicit and explicit tests of memory.},
  author={Gershberg, Felicia B and Shimamura, Arthur P},
  journal={Journal of Experimental Psychology: Learning, Memory, and Cognition},
  volume={20},
  number={6},
  pages={1370},
  year={1994},
  publisher={American Psychological Association}
}

@article{raaijmakers1981search,
  title={Search of associative memory.},
  author={Raaijmakers, Jeroen G and Shiffrin, Richard M},
  journal={Psychological review},
  volume={88},
  number={2},
  pages={93},
  year={1981},
  publisher={American Psychological Association}
}

@article{wiswede2007serial,
  title={Serial position effects in free memory recall—An ERP-study},
  author={Wiswede, Daniel and R{\"u}sseler, Jascha and M{\"u}nte, Thomas F},
  journal={Biological psychology},
  volume={75},
  number={2},
  pages={185--193},
  year={2007},
  publisher={Elsevier}
}

@article{baddeley2009working,
  title={Working memory and binding in sentence recall},
  author={Baddeley, Alan D and Hitch, Graham J and Allen, Richard J},
  journal={Journal of memory and Language},
  volume={61},
  number={3},
  pages={438--456},
  year={2009},
  publisher={Elsevier}
}

@inproceedings{kononenko2015investigating,
  title={Investigating code review quality: Do people and participation matter?},
  author={Kononenko, Oleksii and Baysal, Olga and Guerrouj, Latifa and Cao, Yaxin and Godfrey, Michael W},
  booktitle={2015 IEEE international conference on software maintenance and evolution (ICSME)},
  pages={111--120},
  year={2015},
  organization={IEEE}
}

@article{runeson2009guidelines,
  title={Guidelines for conducting and reporting case study research in software engineering},
  author={Runeson, Per and H{\"o}st, Martin},
  journal={Empirical software engineering},
  volume={14},
  number={2},
  pages={131--164},
  year={2009},
  publisher={Springer}
}

@article{stol2018abc,
  title={The ABC of software engineering research},
  author={Stol, Klaas-Jan and Fitzgerald, Brian},
  journal={ACM Transactions on Software Engineering and Methodology (TOSEM)},
  volume={27},
  number={3},
  pages={1--51},
  year={2018},
  publisher={ACM New York, NY, USA}
}

@inproceedings{beller2014modern,
  title={Modern code reviews in open-source projects: Which problems do they fix?},
  author={Beller, Moritz and Bacchelli, Alberto and Zaidman, Andy and Juergens, Elmar},
  booktitle={Proceedings of the 11th working conference on mining software repositories},
  pages={202--211},
  year={2014}
}

@article{kampenes2007systematic,
  title={A systematic review of effect size in software engineering experiments},
  author={Kampenes, Vigdis By and Dyb{\aa}, Tore and Hannay, Jo E and Sj{\o}berg, Dag IK},
  journal={Information and Software Technology},
  volume={49},
  number={11-12},
  pages={1073--1086},
  year={2007},
  publisher={Elsevier}
}

@article{torkar2021method,
  title={A method to assess and argue for practical significance in software engineering},
  author={Torkar, Richard and Furia, Carlo A and Feldt, Robert and de Oliveira Neto, Francisco Gomes and Gren, Lucas and Lenberg, Per and Ernst, Neil A},
  journal={IEEE Transactions on Software Engineering},
  volume={48},
  number={6},
  pages={2053--2065},
  year={2021},
  publisher={IEEE}
}

@inproceedings{kononenko2016code,
  title={Code review quality: How developers see it},
  author={Kononenko, Oleksii and Baysal, Olga and Godfrey, Michael W},
  booktitle={Proceedings of the 38th international conference on software engineering (ICSE)},
  pages={1028--1038},
  year={2016}
}

@article{lenarduzzi2020some,
  title={Some sonarqube issues have a significant but small effect on faults and changes. a large-scale empirical study},
  author={Lenarduzzi, Valentina and Saarim{\"a}ki, Nyyti and Taibi, Davide},
  journal={Journal of Systems and Software},
  volume={170},
  pages={110750},
  year={2020},
  publisher={Elsevier}
}

@inproceedings{bosu2015characteristics,
  title={Characteristics of useful code reviews: An empirical study at microsoft},
  author={Bosu, Amiangshu and Greiler, Michaela and Bird, Christian},
  booktitle={2015 IEEE/ACM 12th Working Conference on Mining Software Repositories},
  pages={146--156},
  year={2015},
  organization={IEEE}
}

@article{khatoonabadi2023wasted,
  title={On wasted contributions: Understanding the dynamics of contributor-abandoned pull requests--a mixed-methods study of 10 large open-source projects},
  author={Khatoonabadi, SayedHassan and Costa, Diego Elias and Abdalkareem, Rabe and Shihab, Emad},
  journal={ACM Transactions on Software Engineering and Methodology},
  volume={32},
  number={1},
  pages={1--39},
  year={2023},
  publisher={ACM New York, NY}
}

@article{tufano2019empirical,
  title={An empirical study on learning bug-fixing patches in the wild via neural machine translation},
  author={Tufano, Michele and Watson, Cody and Bavota, Gabriele and Penta, Massimiliano Di and White, Martin and Poshyvanyk, Denys},
  journal={ACM Transactions on Software Engineering and Methodology (TOSEM)},
  volume={28},
  number={4},
  pages={1--29},
  year={2019},
  publisher={ACM New York, NY, USA}
}

@inproceedings{vieira2019reports,
  title={From reports to bug-fix commits: A 10 years dataset of bug-fixing activity from 55 apache's open source projects},
  author={Vieira, Renan and da Silva, Ant{\^o}nio and Rocha, Lincoln and Gomes, Jo{\~a}o Paulo},
  booktitle={Proceedings of the Fifteenth International Conference on Predictive Models and Data Analytics in Software Engineering},
  pages={80--89},
  year={2019}
}

@article{herbold2022fine,
  title={A fine-grained data set and analysis of tangling in bug fixing commits},
  author={Herbold, Steffen and Trautsch, Alexander and Ledel, Benjamin and Aghamohammadi, Alireza and Ghaleb, Taher A and Chahal, Kuljit Kaur and Bossenmaier, Tim and Nagaria, Bhaveet and Makedonski, Philip and Ahmadabadi, Matin Nili and others},
  journal={Empirical Software Engineering},
  volume={27},
  number={6},
  pages={125},
  year={2022},
  publisher={Springer}
}

@article{di2017developer,
  title={A developer centered bug prediction model},
  author={Di Nucci, Dario and Palomba, Fabio and De Rosa, Giuseppe and Bavota, Gabriele and Oliveto, Rocco and De Lucia, Andrea},
  journal={IEEE Transactions on Software Engineering},
  volume={44},
  number={1},
  pages={5--24},
  year={2017},
  publisher={IEEE}
}

@inproceedings{nagappan2005use,
  title={Use of relative code churn measures to predict system defect density},
  author={Nagappan, Nachiappan and Ball, Thomas},
  booktitle={Proceedings of the 27th international conference on Software engineering},
  pages={284--292},
  year={2005}
}

@article{da2016framework,
  title={A framework for evaluating the results of the szz approach for identifying bug-introducing changes},
  author={Da Costa, Daniel Alencar and McIntosh, Shane and Shang, Weiyi and Kulesza, Uir{\'a} and Coelho, Roberta and Hassan, Ahmed E},
  journal={IEEE Transactions on Software Engineering},
  volume={43},
  number={7},
  pages={641--657},
  year={2016},
  publisher={IEEE}
}

@inproceedings{wen2019exploring,
  title={Exploring and exploiting the correlations between bug-inducing and bug-fixing commits},
  author={Wen, Ming and Wu, Rongxin and Liu, Yepang and Tian, Yongqiang and Xie, Xuan and Cheung, Shing-Chi and Su, Zhendong},
  booktitle={Proceedings of the 2019 27th ACM Joint Meeting on European Software Engineering Conference and Symposium on the Foundations of Software Engineering},
  pages={326--337},
  year={2019}
}

@article{cosentino2017systematic,
  title={A systematic mapping study of software development with GitHub},
  author={Cosentino, Valerio and Izquierdo, Javier L C{\'a}novas and Cabot, Jordi},
  journal={Ieee access},
  volume={5},
  pages={7173--7192},
  year={2017},
  publisher={IEEE}
}
\end{document}